\documentclass[10pt,conference,letterpaper]{IEEEtran}
\usepackage{graphicx}
\usepackage{balance}  % for  \balance command ON LAST PAGE  (only there!)

\usepackage{amsmath}
\usepackage{color}
\usepackage{algorithm}
\usepackage{algorithmic}
\usepackage{hyperref}
\usepackage{subfigure}
\usepackage{times}
\usepackage{cite}
\usepackage{flushend}

\newtheorem{definition}{Definition}

\title{Digging Deeper into Deep Web Databases by Breaking Through the Top-k Barrier}

\begin{document}

%Actual ICDE format
%\author{%
%{Saravanan Thirumuruganathan{\small $~^{\#1}$}, Nan Zhang{\small $~^{*2}$}, Gautam Das{\small $~^{\#3}$} }%
%% add some space between author names and affils
%\vspace{1.6mm}\\
%\fontsize{10}{10}\selectfont\itshape
%$~^{\#}$ University of Texas at Arlington \\
%\fontsize{9}{9}\selectfont\ttfamily\upshape
%$~^{1}$saravanan.thirumuruganathan@mavs.uta.edu\\
%$~^{3}$gdas@cse.uta.edu%
%% add some space between email and affil
%\vspace{1.2mm}\\
%\fontsize{10}{10}\selectfont\rmfamily\itshape
%$~^{*}$George Washington University\\
%\fontsize{9}{9}\selectfont\ttfamily\upshape
%$~^{2}$nzhang10@gwu.edu
%}

\author{
{Saravanan Thirumuruganathan$^{\ddag}$,
 Nan Zhang$^{\dag\dag}$, 
 Gautam Das$^{\ddag,\dag}$}
\vspace{1.6mm}\\
\fontsize{10}{10}\selectfont\itshape
$^{\ddag}$University of Texas at Arlington $^{\dag\dag}$George Washington University $^{\dag}$Qatar Computing Research Institute  \\
\fontsize{9}{9}\selectfont\ttfamily\upshape
$^{\ddag}$\{{saravanan.thirumuruganathan@mavs,gdas@cse}\}.uta.edu,$^{\dag\dag}$nzhang10@gwu.edu, 
$^{\dag}$gdas@qf.org.qa 
}

\maketitle

\begin{abstract}
A large number of web databases are only accessible through proprietary form-like interfaces which require users to query the system by entering desired values for a few attributes. A key restriction enforced by such an interface is the {\em top-$k$ output constraint} - i.e., when there are a large number of matching tuples, only a few (top-$k$) of them are preferentially selected and returned by the website, often according to a proprietary ranking function. Since most web database owners set $k$ to be a small value, the top-$k$ output constraint prevents many interesting third-party (e.g., mashup) services from being developed over real-world web databases. In this paper we consider the novel problem of ``digging deeper'' into such web databases. Our main contribution is the meta-algorithm GetNext that can retrieve the next ranked tuple from the hidden web database  using only the restrictive interface of a web database without any prior knowledge of its ranking function. This algorithm can then be called iteratively to retrieve as many top ranked tuples as necessary. We develop principled and efficient algorithms that are based on generating and executing multiple reformulated queries and inferring the next ranked tuple from their returned results. We provide theoretical analysis of our algorithms, as well as  extensive experimental results over synthetic and real-world databases that illustrate the effectiveness of our techniques.

\end{abstract}

% NOTE keywords are not used for conference papers so do not populate them

\section{Introduction}
\label{sec:introduction}

\subsection{Problem Motivation} 
Many web databases are ``hidden'' behind (i.e., only accessible via) a restrictive form-like interface which allows a user to form a search query by specifying the desired values for a few attributes; and the system responds by returning a small number of tuples matching the search query.  Almost all such interfaces enforce the {\em top-$k$ constraint} - i.e., when more than $k$ tuples (where $k$ is typically a predetermined small constant) match the user-specified query, only $k$ of them are preferentially selected according to a (often proprietary) ranking function and returned to the user. For example, American Airline's (AA) flight search-by-schedule\footnote{{\em http://www.aa.com/reservation/searchFlightsSubmit.do} By default $k$ = 10. A user may configure $k$ to be as large as 50. No page down is allowed.} has a default value of $k = $ 10. Similarly, Amazon's best sellers list \footnote{\em{http://www.amazon.com/Best-Sellers/zgbs}} for any category only displays the top-100 products.

How to properly set the value of $k$ is an interesting design challenge for a web database owner.  On one hand, the owner may prefer a small $k$ to (1) speed up query processing and shorten the returned webpage, and/or (2) thwart web/tuple scraping. However, in order to accommodate the needs of website users, the value of $k$ should not be too small. Given these two conflicting goals, in practice $k$ is often set to the minimum necessary value, according to the database owner's belief, which provides the user with ``enough'' choices within the returned tuples.  While such a strategy might suffice the simplest use-cases, it often cannot satisfy users with specific needs and also prevents many interesting third-party services from being developed over web databases - e.g.,
\begin{itemize}
\item Consider a third-party service which enables a user to filter query results according to attributes that cannot be specified in the original form-like interface. For example, American Airline's (AA) flight search-by-schedule\footnotemark[1], a top-10 interface, does not allow a user to specify filtering conditions such as finding the top-10 flights with in-flight wifi. If a third-party service wants to provide such a feature, it must somehow ``bypass'' the top-$k$ constraint because otherwise one might not be able to find enough (or any) wifi-equipped flights from the top-10 results. 
\item Consider a web aggregator or a web mashup which joins tuples from multiple hidden web databases and returns the joined results - e.g., a mashup joining Orbitz.com (a hotel booking website) with Tripadvisor.com (a hotel review website) to return the top-$k$ cheapest hotels that have an average review of at least 4 stars. Once again, such a mashup must somehow break the top-$k$ constraint because not enough matching tuples may be discovered from the mere $k$ tuples returned by each web database.
\end{itemize}

To enable these third-party services and many other interesting applications (e.g., data analytics) that are currently disabled/handicapped by the top-$k$ constraint, a trivial solution is for the third-party service provider to negotiate a private agreement with each web database owner in order to establish data-access channels beyond the top-$k$ web interface. Nonetheless, such negotiations are difficult even between large organizations\footnote{\em{http://online.wsj.com/article/SB121755825030403467.html}} due to revenue sharing, security and myriad of other thorny issues - thus making the solution not scalable to a large number of web databases. As such, our focus in this paper is to develop automated third-party algorithms that only use the public interfaces of web databases without requiring any additional cooperation from the database owners.

Another seemingly straightforward solution to the above problems is \emph{crawling} - i.e., the retrieval of \emph{all} tuples in a hidden web database by issuing multiple queries through its web interface \cite{MKK+08,ARJ+07}. Once all tuples are downloaded, they can be treated as a local database to support all of the above applications. Nonetheless, a key pitfall of this solution is its prohibitively high query cost (i.e., the numerous search queries one needs to crawl all tuples from a web database) - which can be simply infeasible for real-world web databases which often impose a per user/IP limit on number of queries one can issue over a given time frame (e.g., Google Search API allows only 100 free queries per user per day).

\subsection{A Novel Problem: Breaking the Top-$k$ Barrier}

Given the pitfalls of crawling, we propose to study in this paper a novel problem of digging deeper into a web database to retrieve (more than $k$) top-ranked tuples which satisfy a user-specified search query - and thereby ``breaking'' the top-$k$ barrier. Specifically, we consider the following fundamental operator:

\vspace{1mm}

\noindent \framebox[\columnwidth]{\parbox{0.9\columnwidth}{\textsc{GetNext}: Given the top-$h$ tuples ($h \geq k$) satisfying a user-specified query, retrieve the next-highest-ranked (i.e., No.$(h+1)$) tuple from the hidden web database by issuing search queries through its public interface, without any knowledge of its ranking function.}}

\vspace{1mm}

One can see that, by calling \textsc{GetNext} iteratively, it is possible to retrieve as many top-ranked tuples as necessary for a user-specified query - thereby enabling both sample applications discussed above without the need of crawling all tuples from the database. Because of the query-number limitations enforced by web databases, an important objective in the design of \textsc{GetNext} is to maintain a small {\em query cost} - a goal shared by most existing studies on exploring hidden web databases (e.g., \cite{DJJ+10, DDM07, JZD11}).
%
%In summary, we have the following problem definition
%
%\smallskip\noindent
%{\bf Breaking the top-$k$ barrier:} {\em For a hidden web database with a top-$k$ interface, given a search query $q$ that can be specified through the input interface, design an automated procedure GetNext that can be iteratively called to retrieve as many top ranked tuples that satisfy $q$ as possible (according to the ranking function used by the hidden web database).}

%\smallskip
%To the best of our knowledge, this novel problem has not been considered in prior research.

%The parameter $h$ also offers its own challenges. For what values of $h$ is it actually feasible to retrieve the top-$h$ tuples? In the extreme case, can one retrieve the entire global order of all tuples in the database, if such an order exists? In this paper we determine necessary and sufficient conditions that will enable solutions for our central problem, and the implications of these conditions on the different problem variants. For example, for certain cases it is simply not possible for any client-side algorithm to determine a total order of the top-$h$ tuples - in this case we develop algorithms that return the most informative partial order.

\subsection{Outline of Technical Results}

To design \textsc{GetNext}, the technical challenge may have subtle differences across various web databases, mainly because of the different ranking functions being used. At one extreme, some websites allow users to choose their own ranking function (from a predetermined set) - e.g., airlines websites allow users to sort by attributes such as by price, departure time, etc. At the other extreme, a website might feature a complex and proprietary {\em query-specific} ranking function (e.g., ``relevance'' of a tuple to a query) that may never be deterministically inferred from other query answers. Other possible ranking functions include a global order that is nevertheless hidden from the input interface - e.g., Amazon uses popularity as the default ranking function but does not allow it to be specified in a search query. For most of the paper, we focus on the case where the ranking function is a query-independent global order of all tuples. The implications of other ranking-function variations on our solutions are discussed separately.

There are two key components of our proposed solution to \textsc{GetNext}: {\em candidate generation} and {\em candidate testing}.

%In candidate generation, the idea is to exploit the information available in the top-$h$ tuples to retrieve a small set of additional tuples from the database that can potentially have the rank $h + 1$. This should be , i.e., that the true $h+1$-th tuple has to belong to this set.  The candidate testing procedure determines the $h+1$-th tuple from the candidate set.

\smallskip\noindent
{\bf Candidate Generation}: Given the top-$h$ tuples, the candidate generation step aims to identify a complete yet small set of tuples that can potentially have the rank $h + 1$. A key observation here is that the problem is equivalent to finding a small set of queries, each of which matches fewer than $k$ tuples in the top-$h$, while together cover the rest of the database. One can see that, since each query in the set returns at least one non-top-$h$ tuples, the No.$(h+1)$ tuple must be returned by at least one query in the set. Based on this key observation, we propose a tuple-chain-construction based technique which further reduces the query cost required for candidate generation significantly.
%
%
%Clearly each such query is guaranteed to retrieve new tuples, and moreover the tuples returned by these constructed queries has to contain the $h+1$-th tuple, since each query returns the top tuples that match its conditions. As we shall show in the paper, it is easy to develop a simple ``baseline'' procedure for candidate generation; however the main challenge is in developing a candidate generation procedure with a small query cost. One of our main algorithmic contributions is an innovative candidate generation procedure that works by returning a set of {\em chains of tuples} instead of individual tuples - where each chain represents a set of discovered tuples that are known to be consecutively ordered. The candidates then become the highest ranked nodes of each chain.  The testing phase then has to only focus on testing a subset of the highest ranked node of each chain, and when one of them passes the test, all the nodes in that chain form the next ranked tuples. By this mechanism, it is possible to discover multiple consecutively ordered tuples in a single iteration.

\smallskip\noindent
{\bf Candidate Testing:} Since the task is now reduced to testing which candidate is the No.$(h+1)$ tuple, the key enabling question becomes how to perform pairwise rank-comparison between two tuples. Interestingly, for certain pairs of tuples, the comparison may be done with a single query to the hidden database. Specifically, consider issuing the most specific query that matches both tuples. If both are returned, then the result reveals their order. If only one is returned, then it must have a higher rank. The challenge, however, is in the worst-case scenario where neither is returned. In the paper, we start by resolving this scenario with a baseline approach that requires $2^m$ queries, where $m$ is the number of attributes. Then, we propose two ideas - one connects with the well-studied problem of {\em minimal infrequent itemsets} mining \textcolor{red}{\cite{}}, and the other is a heuristic of query-result inference - which significantly reduce the query cost for candidate testing.

\subsection{Summary of Contributions}
In summary, the main contributions of this paper are:

\begin{itemize}

\item	 We introduce the novel problem of breaking the top-$k$ barrier of a hidden web database to retrieve top ranked tuples that match a user query. We consider several variants of the problem, and study necessary and sufficient conditions under which this problem can be solved.

\item	 We propose BEYOND-$h$-GETNEXT and ORDERED-GETNEXT, two algorithms that iteratively uses the two fundamental operations, candidate generation and candidate testing, to retrieve the next-highest-ranked tuple. While BEYOND-$h$-GETNEXT guarantees the correct retrieval of next ranked tuple\footnote{if such an order can be uniquely determined from the top-$k$ interface.}, ORDERED-GETNEXT further uses an effective heuristic of query-result inference to significantly reduce the query cost in practice without sacrificing correctness.

%\item	Our main idea for candidate generation hinges upon the fact that the returned results of a query gives ordering information for $k$ tuples, which allows us to retrieve multiple consecutive ranked tuples in a single invocation.
%
%\item	Our main ideas for candidate testing avoids the need to perform a complete crawl of the database. We develop the procedure BEYOND-$h$-TEST whose performance is related to the problem of generating and counting the number of infrequent minimal infrequent itemsets in a database. We also develop the procedure ORDERED-TEST, based on heuristics that order the execution of the generated queries in such a way that early terminating of the procedure can be achieved.

\item Our contributions also include a careful theoretical analysis of BEYOND-$h$-GETNEXT and ORDERED-GETNEXT, as well as a through experimental evaluation over both synthetic datasets and real-world websites.

\end{itemize}

The rest of the paper is organized as follows. In $\S$2, we discuss preliminaries - e.g., the models of hidden web databases and their ranking functions. $\S$3 defines the problem of breaking the top-$k$ barrier and outlines our proposed solution that uses \textsc{GetNext}. $\S$4 and $\S$5 detail the two main parts of our algorithm, candidate generation and candidate testing, respectively. In $\S$6, we discuss extensions to the algorithms to handle special cases. $\S$7 describes a detailed set of experiments over real-world datasets. $\S$8 discusses related work, followed by the conclusion in $\S$9.

\section{Preliminaries}
\label{sec:preliminaries}

In this section, we introduce a model for hidden databases and describe the different types of ranking functions used commonly in hidden databases.

\subsection{Model of Hidden Databases}
\label{subsec:hiddenDBModel}

Consider a hidden database $D$ with $n$ tuples and $m$ input attributes $A_1, A_2, \ldots, A_m$. Given a tuple $t$ and attribute $A_i$, let $t[A_i]$ be the value of $A_i$ in $t$. Let $Dom(A_i)$ be the domain of $A_i$. For the purpose of this paper, we restrict our attention to categorical attributes and assume the appropriate discretization of numeric ones. We also consider all tuples distinct and without null values. Let $f(.)$ be the ranking function which takes a tuple and a query as input and outputs an integer between $1$ and $n$. Without loss of generality, we assume the output of $f(.)$ to be unique for each tuple.

A user can query the system by specifying the desired values for a subset of $A_1, \ldots, A_m$. Thus, a user query $q$ is of the form \texttt{SELECT * FROM $D$ WHERE} $A_{i_1} = v_{i_1} \& \ldots \& A_{i_s} = v_{i_s}$, where $\{i_1, \ldots, i_s\} \subseteq [1,m]$. and $v_{i_j} \in Dom(A_{i_j})$. The set of tuples matching query $q$ is denoted as $Sel(q)$. If $|Sel(q)| > k$, an $overflow$ occurs and only the top-$k$ results are returned, along an overflow flag indicating that more tuples matching the query cannot be returned. If $|Sel(q)|=0$, then an {\em underflow} occurs as no tuples match the query. Otherwise, i.e., when $|Sel(q)| \in [1, k]$, we say that $q$ is valid. For the purpose of this paper, we make the realistic assumption that $k > 1$.

For the purpose of our paper, we assume that the interface only displays the top-$k$ results and does not allow users to extract additional results by scrolling through the results. The only way to get additional results is to reformulate the input query. This is a reasonable assumption as many real world hidden web databases such as Yahoo!~Autos limit the maximum number of page turns a user can perform. 

\subsection{Model of Ranking Function}
\label{subsec:rankingFunctionModel}
There are two broad categories of ranking functions: {\em static} and {\em query-dependent}.

\begin{itemize}
\item A ranking function $f(.)$ is \textit{static} if for a given tuple $t$, $f(q, t)$ is constant for all queries $q$ - i.e., the rank of a tuple is independent of the query being issued. An example in practice is the ``sort by price'' used by Yahoo!~Autos. Note that the input tuple may feature not only $A_1, \ldots, A_m$ but also the non-input-specifiable attributes (e.g., ``popularity'' as discussed in $\S$1).  
\item A ranking function is \textit{query-dependent} if, for a given $t$, $f(q, t)$ varies for different queries $q$. An example of such a ranking function occurs in a fuzzy-matching scenario where all tuples are ordered according to the number of attribute matches between the query and each tuple.
\end{itemize}

As discussed in $\S$1, we focus on static ranking functions in this paper. The reason for doing so is simple - if the ranking function is query dependent, no mechanism can be used to fetch the next ranked tuple. To understand why, note that in order to get tuples beyond top-$k$, it is necessary to reformulate the query. But this has the side effect of arbitrarily changing the ranking of tuples. Hence, with a query-dependent ranking function, no mechanism can guarantee the discovery of tuples with rank greater than $k$ for a given query.

%As discussed in $\S$1, we focus on static ranking functions in this paper. The reason for doing so is simple. We assume that there exists a global order between the top ranked tuples. Consider an arbitrary query-dependent ranking function $f(q,t)$. To correctly retrieve the top ranked tuples, the function $f(q,t)$ must return the tuples in the same global order for any query $q$. In other words, if for any two tuples $u$ and $v$ and two arbitrary queries $q$ and $q'$, if $f(q,u) > f(q,v)$ and $f(q',u) < f(q',v)$, then no such global order can exist. However, if there exist a global order of tuples, then the order holds irrespective of the query $q$ which makes the ranking function $f(q,t)$ to be {\em static}.

For the purpose of this paper, we conservatively assume that the ranking function is unbeknown to our algorithm. If the ranking function is known and is based on the attributes returned by the hidden web interface (such as sort by price), it is possible to leverage this information to design algorithms with significantly less query cost. We further discuss this variant in $\S$5. In addition, we assume that it is possible to infer a unique global order of the top-ranked tuples to be extracted from the web interface. If such an order cannot be inferred from the interface, one of the possible partial orders would be returned, as we shall explain in $\S$6.

\vspace{1mm}

\noindent \framebox[\columnwidth]{\parbox{0.9\columnwidth}{\textbf{Running Example: } Table~\ref{tbl:runningExample} shows a simple table which we shall use as running example throughout this paper. There are $m$ = 5 Boolean attributes and $n$ = 7 tuples which are ranked in the order given in the table. i.e., $t_1$ is the highest ranked tuple.}}

\vspace{1mm}

\begin{table} 
\label{tbl:runningExample}
\centering
\caption{Input Table, $D$}
\begin{tabular}{cccccc} \hline
& $A_1$ & $A_2$ & $A_3$ & $A_4$ & $A_5$ \\ \hline 
$t_1$ & 0 & 0 & 0 & 0 & 1 \\ \hline
$t_2$ & 0 & 0 & 0 & 1 & 1 \\ \hline
$t_3$ & 0 & 0 & 1 & 0 & 1 \\ \hline
$t_4$ & 0 & 1 & 1 & 1 & 1 \\ \hline
$t_5$ & 1 & 1 & 1 & 0 & 1 \\ \hline
$t_6$ & 1 & 1 & 1 & 1 & 1 \\ \hline
$t_7$ & 0 & 0 & 0 & 0 & 0 \\ \hline
\end{tabular}
\end{table}

\section{Overview of GetNext}
\label{sec:problemDefAndIntution}
In this section, we first discuss the technical challenges of \textsc{GetNext}, and then outline the structure of our proposed two-step solution - the details of each step shall then be developed in the next two sections, respectively.

%\subsection{Problem Definition}
%\label{subsec:problemDef}
%Recall from Section~\ref{sec:preliminaries} that our aim to design \textsc{GetNext} operator that enables the retrieval of $h+1$-th tuple for any $h > k$ from a hidden web database $D$ which only exposes the top-$k$ ranked tuples in the query answers it returns through the restrictive web interface. The hidden database does not disclose the ranking function it uses. The \textsc{GetNext} operator issues (a carefully selected set of) multiple queries and then infers the $h+1$-th tuple based on the ordering of returned tuples exposed by multiple query answers. Formally, the problem can be stated as follows.
%
%\vspace{2mm} \noindent \textbf{Problem Statement :} \textit{Given a hidden web database with a form based input interface that restricts user access to top-$k$ tuples ordered by an unknown static ranking function, discover the $h+1$ ranked tuple for any $h > k$ while minimizing the query cost.}

\subsection{Technical Challenges}
\label{subsec:basicIdeas}

To illustrate the main technical challenges, we consider a fundamental question: Given two tuples $t$ and $t^\prime$, how can we determine which one ranks higher? We start with a straightforward comparison - i.e., when $t$ and $t^\prime$ match the same query $q$ which returns at least one of the two tuples:
\begin{itemize}
\item if $q$ returns $t$ but not $t^\prime$, then $t$ is ranked higher,
\item if $q$ returns $t^\prime$ but not $t$, then $t^\prime$ is ranked higher, or
\item if $q$ returns both, then we can make the comparison based on the returned order.
\end{itemize}
In this case, we call two tuples {\em directly comparable}, with the higher-ranked tuple {\em directly dominating} the other one - i.e.,

\begin{definition}
{\bf [Domination] \label{def:domination}}
A tuple $t$ is said to \textbf{directly dominate} another tuple $t^\prime$, i.e., $t \succ t^\prime$, if and only if $t$ and $t^\prime$ are directly comparable and $t$ ranks higher than $t^\prime$.
\end{definition}

A tuple can dominate another tuple directly or indirectly. Suppose tuple $t \succ u$ and $u \succ v$. Even if $t$ and $v$ are not directly comparable, we can infer that $t$ indirectly dominates $v$. By default, we use the term domination to refer to direct domination. 

For example, consider the running example with a top-2 interface. We can observe that $t_1$ and $t_3$ are directly comparable using the query \texttt{$q_1$: SELECT * FROM D WHERE $A_1=0$ AND $A_2=0$ AND $A_4=0$ AND $A_5=1$} with $t_1$ ranked higher than $t_3$. Similarly, tuples $t_2$ and $t_3$ are directly comparable using the query \texttt{$q_2$: SELECT * FROM D WHERE $A_1=0$ AND $A_2=0$ AND $A_5=1$}. The result includes $t_2$ but not $t_3$ - i.e., $t_2$ ranks higher.

A key observation here is that if two tuples are directly comparable, then we need only one query to determine their domination relationship: the {\em most specific} query which matches both tuples - i.e., the query which contains one predicate for each attribute on which both tuples share the same value. To understand why, note that if this query cannot return at least one of the two tuples, then no other query can - i.e., the two tuples are not directly comparable. For the running example, both $q_1$ and $q_2$ shown above are the most specific queries matching the two corresponding tuples.

While the possibility of direct comparison shows promises for ranking tuples in the database, it also illustrates the key technical challenge for \textsc{GetNext}: not every pair of tuples are directly comparable with each other - e.g., neither $t_6$ nor $t_7$ in the running example can be returned by the most specific query that matches both of them (i.e., SELECT *). 

In this case, the comparison of the two tuples requires one to identify a ``bridge'' of tuples between them - e.g., $t \succ t_x \succ t^\prime$ for comparing $t$ with $t^\prime$. The problem, however, is it is unclear how one can find the bridging tuples without actually crawling all tuples from the database and incurring a prohibitively high query cost. In the next subsection, we outline the structure of our proposed solution to address this challenge.

%A graphical way to represent the comparability of tuples is by using a directed acyclic graph (DAG). The nodes correspond to tuples and a directed edge exists between two nodes $u$ and $v$ if the respective tuples are directly comparable and $u$ has a higher rank than $v$. Even if two tuples are directly comparable, their relative rank can still be inferred if a directed path exists between the corresponding nodes in the graph. In this case, the tuple corresponding to the start node has a higher rank than the one for end node. The DAG can be filled by performing pairwise comparison between the respective tuples. If the tuples are not directly comparable, then no edges exist between them. The simplest DAG that can be constructed is that of a chain where each element is atleast directly comparable to its neighbors.

\subsection{Outline of Our Proposed Solution}
Our proposed solution for \textsc{GetNext} is a two-step process: 

\begin{itemize}
\item {\em Candidate Generation:} In this step, we identify a small set of {\em candidate tuples} which are guaranteed to contain the No.~$h+1$ tuple. If the output set has a size of 1, then we can directly output the No.~$h+1$ tuple. Otherwise, we call the following candidate testing step.  $\S$4 describes our design for candidate generation.

\item {\em Candidate Testing:} In this step, we take the set of candidate tuples as input and compare between them to determine which tuple is indeed the No.~$h+1$. $\S$5 describes our design for candidate testing. 
\end{itemize}

%
%\begin{algorithm}[!htb]
%\caption{\textbf{ TOP-\lowercase{h}-META}}
%\begin{algorithmic}[1]
%\label{alg:metaAlgo}
%\STATE \textbf{Input parameters :} $h$
%\FOR {$i = 1$ to $h - 1$}
%\STATE \textbf{Step Generation: } Generete candidates for tuple $t_{i+1}$ based on knowledge of $t_1, \ldots, t_i$.
%\STATE \textbf{Step Testing: } Given the candidate set, determine and return $t_{i+1}$.
%\ENDFOR
%\end{algorithmic}
%\end{algorithm}

\section{Candidate Generation}
\label{sec:candidateGeneration}

We now consider the detailed design of candidate generation. Given the current set of top ranked tuples, the candidate generation step is supposed to produce a set of candidate tuples, one of which is guaranteed to be the next ranked tuple. The determination of the exact next-ranked tuple from the candidate set is done using the candidate testing oracle described in Section V. In this section, we first describe a baseline approach for candidate generation, and then introduce a more efficient algorithm using a notion of directed acyclic graphs (DAG) of tuples. The DAG based algorithm exploits the ordering information provided by query answers to potentially complete multiple rounds of candidate generation in a single iteration (i.e., it may answer multiple consecutive \textsc{GetNext} calls without additional query cost). Recall from Section II that we make the realistic assumption of $k > 1$.

\subsection{Baseline Approach}
The essence of candidate generation can be stated as follows. Given the top-$h$ tuples, candidate generation needs to identify a set of queries that is guaranteed to ``cover'' (i.e., return) the next-ranked (i.e., No.$(h+1)$) tuple. One can see that such a set of queries must together match all possible tuples in the database - in order to ensure that no other tuple has a higher rank than the next-ranked tuple being covered.

We start by considering a simple baseline approach as follows: First, find a set of attributes $\mathcal{A}$ such that if we partition the top-$h$ tuples based on their value combinations for attributes in $\mathcal{A}$, then each partition contains fewer than $k$ elements.  Since each tuple is unique, such an $\mathcal{A}$ already exists. After finding $\mathcal{A} = \{A_{i_1}, \ldots, A_{i_j}\}$, we construct queries of the form $q_i$: \texttt{SELECT * FROM D WHERE} $A_{i_1} = v_{i_1}$ AND $\cdots$ AND $A_{i_j} = v_{i_j}$ for all possible value combinations of $v_{i_1} \in Dom(A_{i_1}), \ldots, v_{i_j} \in Dom(A_{i_j})$, and execute all such queries. One can see that these queries completely cover the database domain and thus return a candidate set for the No.$(h+1)$ tuple. To understand why, note that the No.$(h+1)$ tuple must be returned by one of the queries issued, because otherwise the query which matches the No.$(h+1)$ tuple must return a tuple that directly dominates the No.$(h+1)$ tuple.

\vspace{1mm}

\noindent\textbf{Example 1:} Given the top-$3$ tuples in the running example, suppose we want to retrieve the next ranked tuple. We identify an attribute, say $A_3$ (or $A_4$), such that the number of tuples having the values $0$ and $1$ are less than $k=3$. We execute two queries by augmenting $q$ - specifically, $q_1$: \texttt{SELECT * FROM D WHERE $A_3=0$} returns new tuples $\{t_7\}$  and $q_2$: \texttt{SELECT * FROM D WHERE $A_3=1$} returns new tuples $\{ t_4,t_5\}$. The candidate set for 4-th ranked tuple is the set $\{t_4,t_5,t_7\}$. If we want to retrieve the 5-th ranked tuple, we can choose any of the attributes $A_2,A_3$ or $A_4$ to partition the top-4 tuples.

\vspace{1mm}

\noindent\textbf{Analysis:} The number of queries executed to identify the candidate set depend on the domain value of the attribute(s) selected. Given an attribute set $\mathcal{A}$, the number of queries executed is $\prod_{A \in \mathcal{A}} |Dom(A)|$.

\begin{figure}[t]
	%\centerline{\includegraphics[scale=0.4,trim = 10mm 15mm 10mm 10mm, clip]{figs/dag.eps}}
	\centerline{\includegraphics[scale=0.4]{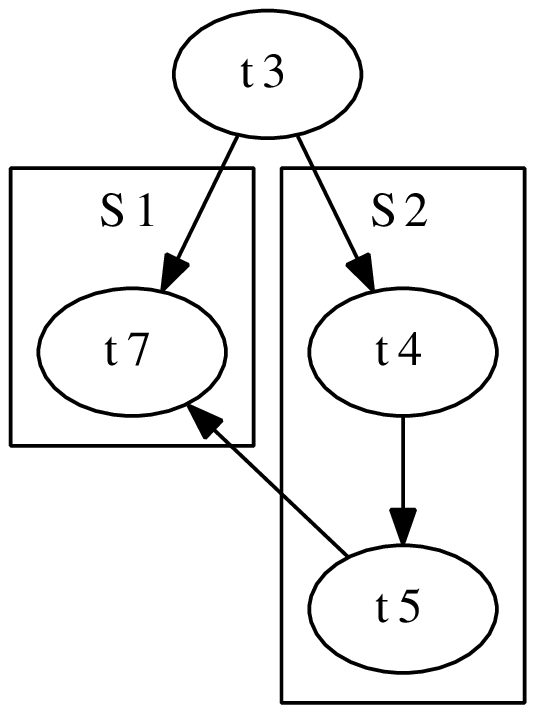}}
    \caption{DAG used in Examples 1 and 2}
    \label{fig:DAG}
	\vspace{-5mm}
\end{figure}

\subsection{DAG based Approach}
In this subsection, we develop a DAG-based algorithm which leverages the order information provided in the query results to further reduce the number of returned candidate tuples, and to identify the candidate sets for multiple next-ranked tuples at a single iteration. In other words, our DAG based approach retrieves the candidate sets for as many next ranked tuples as possible so that subsequent {\em GetNext} do not incur any additional query cost.

%Consider an extreme example where the baseline approach performs poorly. Suppose we partitioned the top-$h$ tuples using a boolean attribute $A$. There are two augmented queries $p$ and $q$ based on $A$. Let the ordered unseen tuples from $p$ results be $u_1$ to $u_x$ while that from $q$ be $v_1$ to $v_y$. Suppose, $u_x$ dominated $v_1$. In this case, we can see that the tuples $u_1$ to $u_x$ are the next consecutive ranked tuples. However, the baseline approach cannot make this inference.

The data structure used in our approach is a directed acyclic graph (DAG) called the {\em dominance directed graph}. Each node in the DAG correspond to a tuple and a directed edge exist from node $u$ to node $v$ if $u$ dominates $v$. Given the result of any query $q$, we can form an DAG from it results. If the query returned $|q|$ tuples, then the DAG would have at most ${|q| \choose 2}$ edges and an linear chain of $|q|$ tuples as a subgraph. An example of the DAG formed from queries $q_1$ and $q_2$ from Example 1 is in Figure~\ref{fig:DAG}. Given a set of queries $q_i$, we can form a set of linear chains from their results. Let $S_i$ denote the $i$-th linear chain and $S$ be the set of all linear chains. The notation $head(S_i)$ returns the tuple with highest rank in $S_i$ while $head(S)$ returns the set of highest ranked tuples in each chain.

The primary aim of this approach is to identify a linear of chain of consecutively ranked tuples, if any. If such a chain exists, then the tuples from the chain can be returned for the subsequent {\em GetNext} calls without additional query cost. We use two observations to extract this chain. First, the only tuples that can dominate the candidates for $t_{h+1}$ are the ones in the top-$h$. Second, since the database has a fixed (but hidden) global order of all tuples, there always exists a dominance relationship (i.e., direct comparison) between the tuples with rank $h$ and $h+1$. If not, the ranks of these two tuples can be flipped without violating any other relative rankings. 

To see how these observations are useful, consider the augmented queries from the baseline approach. Each such query $q_i$ results in a linear chain $S_i$. We can see that $head(S_i)$ dominates other tuples from $S_i$. Hence, $head(S_i)$ is the only tuple from $S_i$ that needs to be added to candidate set. Since tuples $t_h$ and $t_{h+1}$ must be directly comparable, we need to consider only the head of each linear chain and compare it with tuple $t_h$. 

The overview of the algorithm is as follows. We have a list of linear chains (from augmented queries of prior {\em GetNext} invocations) and the linear chain, say $S_i$, from which tuple $t_h$ was extracted. We perform pairwise comparison between tuples from different linear chains. An edge is added from node $u$ to node $v$, if they are directly comparable and $u$ ranks higher than $v$. Then we compare the tuple $t_h$ with the head of each chain {\em except} $S_i$. If {\em none} of the heads are directly comparable with $t_h$, then we can assign $head(S_i)$ to be the next ranked tuple without even performing candidate testing. This is possible due to the fact that consecutively ranked tuples are always comparable. If some of them are comparable with $t_h$, only these form the candidate set for $t_{h+1}$. The candidate tuples are then compared pairwise with each other to identify non dominated tuples. The domination can be either direct or indirect. It is easy to see that tuple $t_{h+1}$ is guaranteed to be among the non dominated tuples that are also comparable to tuple $t_h$.

If there are multiple candidate tuples for $t_{h+1}$, then the candidate testing oracle must be invoked. If not, we are guaranteed that the only candidate tuple must have rank $h+1$. The candidate tuple is then removed from its linear chain and the process is continued till the number of candidates for the next ranked tuple is more than 1. This can potentially result in multiple consecutive next ranked tuples to be retrieved. 

\vspace{1mm}

\noindent\textbf{Example 2:} Consider the same setting as Example 1. We wish to extract 4-th ranked tuple from a top-3 interface. Using attribute $A_3$, we construct two augmented queries $q_1$ and $q_2$ resulting in two linear chains $S_1$ and $S_2$. The last tuple $t_3$ belonged to linear chain $S_2$. The resulting DAG can be seen from Figure ~\ref{fig:DAG}. Both the tuples $t_7$ and $t_4$ are comparable with $t_3$ and do not dominate each other. However, $t_7$ is indirectly dominated by $t_4$ through $t_5$. Hence we can immediately declare $t_4$ as the 4-th ranked tuple. Since $t_5$ also dominates $t_7$, it is identified as the 5-th ranked tuple. Note that in both the cases, no calls were made to the candidate testing section. Additionally, we identified two consecutively ranked tuples in a single invocation of {\em GetNext}.

\vspace{1mm}

\textbf{Analysis :} At each iteration, let the number of linear chains be $l$. The query cost for pairwise comparison of tuples between chains is $\prod_{i=1}^{l} |S_i|$. We also require an addition $l$ queries to compare tuple $t_h$ with the heads of each chain. Thus, the algorithm requires at most $l + \prod_{i=1}^{l} |S_i|$ in any iteration. Note that subsequent iterations do need any additional queries till one of the chains is completely consumed as the comparison information between tuples has already been identified.

\section{Candidate Testing}
\label{sec:candidateTesting}

In this section, we consider the candidate testing problem - i.e., based on prior knowledge of the top-$h$ ranked tuples $t_1, \ldots, t_h$, what queries does one need to issue to the hidden database in order to test whether a given tuple $t$ has  rank $h+1$ ? We start with two baseline approaches which can require prohibitively high query costs in practice, and then present our two ideas for improving their efficiency: (1) a reduction to beyond-$h$ minimal queries - which significantly reduces both worst- and average-case query costs, and (2) a heuristic query ordering - which further reduces the query cost in practice. It must be noted that if the ranking function is known and based on the attributes returned by the hidden database (e.g. sort by price), then the next ranked tuple can be directly identified from the candidate tuples without an explicit candidate testing phase or querying the hidden database for comparison.

\subsection{Baseline Approaches}

%We start by discussing the requirement of rank testing %Like in the previous section, we focus on the case where all attributes are Boolean - with extensions to generic cases discussed in Section~\ref{sec:ext}.
To prove that $t$ indeed has rank $h+1$, we have to ensure that no tuple in the database, other than the top-$h$ ones, dominates $t$.  A seemingly straightforward baseline approach is then to first crawl all other tuples from the database, and then compare each of them with $t$ to identify any dominance relationship. The problem with this approach, however, is that the crawling step requires at least $n/k$ queries - where $n$ is the number of tuples in the database and $k$ is as in the top-$k$ interface - because each query returns at most $k$ tuples. Most common hidden web databases routinely have hundreds of thousands of tuples with a relatively small value of $k$, resulting in a prohibitive query cost to test a single tuple.

We now consider another baseline which is enabled by the following observation: according to the definition of dominance relationship shown in $\S$3, the only queries which may ``reveal'' a tuple dominating $t$ are those that actually match $t$ - i.e., queries of the form
\begin{align}
q: \mbox{SELECT * FROM D WHERE } &A_{i_1} = t[A_{i_1}] \mbox{ AND } \cdots \nonumber\\
\mbox{AND } &A_{i_r} = t[A_{i_r}] \label{equ:qdf}
\end{align}
where $\{i_1, \ldots, i_r\} \subseteq \{1, \ldots, m\}$ (recall that $m$ is the number of attributes). Specifically, $t$ has rank $h+1$ if and only if every query of the form (\ref{equ:qdf}) either returns $t$ as the highest-ranked non-top-$h$ tuple, or returns only tuples in the top-$h$.

Thus, our second baseline is to issue all queries matching $t$. One can see that the query cost for the second baseline is $({m \atop 0}) + \cdots + ({m \atop m}) = 2^m$. While this number is often much smaller than $n/k$ for a practical hidden database (because there are usually only a few, e.g., 5 or 10, attributes that can be specified on the input web interface), issuing $2^m$ queries for each candidate tuple may still lead to an extremely high query cost. In the following two subsections, we develop our two ideas for reducing query cost respectively.

\subsection{Beyond-$h$ Minimal Queries} \label{sec:msc}

Our first idea is to reduce the space of queries required for rank testing from all queries which match $t$ (i.e., of the form in (\ref{equ:qdf})) to a much smaller subset which we refer to as the {\em beyond-$h$ minimal queries}. In the following, we first define beyond-$h$ minimal queries and show the completeness of such queries - i.e., issuing them suffices for rank testing. Then, we describe a (somewhat surprising) mapping of beyond-$h$ minimal queries to finding minimal infrequent itemsets - a problem that has been extensively studied in the database and data mining communities (e.g., see survey in \cite{HCXY07}). Finally, we leverage the existing results on minimal infrequent itemsets to derive an upper bound on the number of beyond-$h$ queries.

\vspace{2mm} \noindent {\bf Definition and Completeness:} For any query $q$ which matches $t$, we use $S(q)$ to represent the {\em companion attribute set} of the query - i.e., the set of attributes involved in the query. For example, $S(q) = \{A_{i_1}, \ldots, A_{i_r}\}$ for $q$ in (\ref{equ:qdf}). Then, we call $q$ a {\em beyond-$h$ minimal query} if and only if it satisfies both of the following two conditions:
\begin{itemize}
\item $q$ must return at least one non-top-$h$ tuples - i.e., $q$ must match fewer than $k$ tuples in $t_1, \ldots, t_h$
\item any query $q^\prime$ which matches $t$ and has $S(q^\prime) \subset S(q)$ must only return top-$h$ tuples - i.e., $q^\prime$ must match at least $k$ tuples in $t_1, \ldots, t_h$.
\end{itemize}

One can see from the definition that, as the name suggests, $q$ is a ``minimal'' query which returns any tuple beyond the top-$h$. We now explain why issuing only beyond-$h$ minimal queries suffices for rank testing. Consider the testing of whether $t$ is the tuple with rank $h+1$. A key observation here is that any query $q_0$ which matches $t$ but is not a beyond-$h$ minimal query must satisfy one of the following two conditions:
\begin{itemize}
\item If $q_0$ matches at least $k$ tuples in $t_1, \ldots, t_h$, then one can already infer the answer to $q_0$ from the knowledge of $t_1, \ldots, t_h$ - i.e., $q_0$ is useless for rank testing.
\item If $q_0$ matches fewer than $k$ tuples in $t_1, \ldots, t_h$ but is not a beyond-$h$ minimal query, then there must exist a beyond-$h$ minimal query $q^\prime_0$ such that $S(q^\prime_0) \subset S(q_0)$. If $q^\prime_0$ returns $t$ as the top-ranked tuple besides top-$h$, then we are already certain that no non-top-$h$ tuple matching $q_0$ can outrank $t$. Otherwise, we are already certain that $t$ cannot have rank $h+1$ - i.e., in either case, we do not need to issue $q_0$.
\end{itemize}

\textbf{Example :} Considering the running example from Table~\ref{tbl:runningExample}, we can see that $A_3=1$ and $A_4=1$ are two examples of beyond-$h$ queries for $t_4$.

\vspace{2mm} \noindent {\bf Mapping:} We now show that the problem of finding all beyond-$h$ minimal queries is equivalent to finding all minimal infrequent itemsets over a transactional database. To understand why, consider the following procedure which maps the top-$h$ tuples to $h$ transactions. We first map each attribute $A_j$ ($j \in [1, m]$) to an item $s_j$. Then, for each tuple $t_i$ ($i \in [1, h]$), we map it to a transaction $r_i$ by including in $r_i$ all items corresponding to the attributes on which $t_i$ and the testing tuple $t$ share the same value - i.e.,
\begin{align}
r_i = \{s_j | t_i[A_j] = t[A_j]\}.
\end{align}

We can see that, with this mapping, the companion attribute set of each beyond-$h$ minimal query $q$, i.e., $S(q)$, becomes a minimal infrequent itemset over the $h$ transactions, with the frequency threshold being $k / h$. This observation can be readily made from the definition of beyond-$h$ minimal queries: Since such a query must match fewer than k tuples in $t_1, \ldots, t_h$, $S(q)$ is infrequent given the threshold of $k/h$. Since no subset of $S(q)$ can match fewer than $k$ tuples in top-$h$, $S(q)$ must be minimally infrequent. One can see that the inverse also holds - i.e., there is a one-one mapping between $S(q)$ and a minimal infrequent itemset.

\noindent \textbf{Example :} Suppose we have extracted the top three tuples and want to determine if tuple $t_4$ is indeed the 4-th ranked tuple. We first map tuples $t_1,t_2,t_3$ to transactions as $r_1=\{A_1=0,A_5=1\}, r_2=\{A_1=0, A_4=1,A_5=1\} \text{ and } r_3=\{A_1=0,A_3=1,A_5=1\}$. The threshold is $\frac{3}{3}=1$. The infrequent itemsets are $A_3=1$ and $A_4=1$ which correspond to beyond-$h$ queries for $t_4$. Also, the number of beyond-$h$ queries is dramatically smaller than the $2^5$ queries needed in the previous approach.

While (as we shall show below) the mapping enables us to derive an upper bound on the number of beyond-$h$ minimal queries, we would like to remark here two major differences between our problem and the traditional problem of finding minimal infrequent itemsets.

First, even though finding all minimal infrequent itemsets is known to be \#P-complete, the time complexity is {\em not} really a concern for our problem because our input size $m$ - i.e., the number of attributes - is usually much smaller than the number of items in a transactional database. As such, we could simply enumerate all $2^m$ possible itemsets (and find the minimal infrequent ones) without causing significant overhead. What is a major concern for us, however, is the number of minimal infrequent itemsets because it translates to the number of queries we have to issue through the web interface - a costly and time-consuming process.

Second, our frequency threshold, i.e., $k / h$, is generally much larger than the threshold traditionally considered for minimal infrequent itemsets. As we mentioned in $\S$1, even an $h = 2k$ may bear significant interest as third-party analyzers are most likely interested in those highly ranked, albeit outside top-$k$, tuples. As we shall show below, this unusually high threshold enables us to improve the upper bound on the number of beyond-$h$ minimal queries when $h$ is small.

\vspace{2mm} \noindent {\bf Upper Bound:} First, according to the existing results on the number of minimal infrequent itemsets, that the number of beyond-$h$ minimal queries can be bounded by $\left(m \atop {m/2}\right)$. We now show that when $h$ is small, specifically $h \leq m/2 + k - 1$, the number of beyond-$h$ minimal query $q$ has another upper bound of $\left(m \atop {h-k+1}\right)$. 

An important observation here is that the number of predicates in a beyond-$h$ minimal query, say $q$, is at most $h - k + 1$. To understand why, consider a query-construction process in which we start with the SELECT * query, and then gradually add into it one conjunctive predicate in $q$ (i.e., one attribute in $S(q)$) at a time, until the query matches fewer than $k$ tuples in the top-$h$. One can see that each predicate being added, say $A_i = t[A_i]$, must remove at least one top-$h$ tuple from the set of tuples matching the previous query, because otherwise one can always remove $A_i = t[A_i]$ from $q$ without changing the answer to $q$ - contradicting the fact that $q$ is beyond-$h$ minimal. As such, once $h-k+1$ predicates are added to the query, the number of top-$h$ tuples matching the query must drop to below $k$ - i.e., $S(q)$ contains at most $h - k + 1$ attributes. Again, since all beyond-$h$ minimal queries forms an anti-chain, the number of them is at most $\left(m \atop {h-k+1}\right)$ when each beyond-$h$ minimal query contains at most $h - k + 1$ predicates and $h - k + 1 \leq m/2$.

In summary, we have the following theorem:

\newtheorem{theorem}{Theorem}
\begin{theorem} \label{thm:exb}
Given the top-$h$ tuples, the maximum number of queries one needs to issue for testing whether a tuple has rank $h+1$ over a database of $m$ attributes and $n$ tuples, $c(n, m, h+1)$, satisfies
\begin{align}
c(n, m, h+1) \leq \left(m \atop {\min(h-k+1, m/2)}\right).
\end{align}
\end{theorem}

Using the fact that $\sum_{i=1}^m {m \choose i} = 2^m$, we can show a tighter upper bound for the number of beyond-$h$ queries as $\frac{2^m}{(m+1)}$, resulting in substantial reduction in query cost over the baseline approaches. 

\subsection{Query Ordering}
Our next idea to reduce query cost that works very well in practical hidden databases is a heuristic - query ordering. Recall that {\em beyond-$h$ query} is a minimal query that returns at least one non-top-$h$ tuple. Given a candidate tuple $t$, if all its corresponding beyond-$h$ minimal queries returns $t$ as the highest ranked non-top-$h$ tuple, then we can conclude that no other tuple dominates $t$ and hence $t$ has rank $h+1$. Note that to make this conclusion, it is mandatory to execute all the beyond-$h$ queries. 

The key idea in query ordering is that of {\em elimination}. If we can eliminate all but one tuple from the candidate set, then the remaining tuple has to be the next ranked tuple and we can make that conclusion even without executing any of the beyond-$h$ queries for it. This is due to the fact that the candidate generation step produces a set of tuples one of which is guaranteed to be in the next ranked tuple. The query ordering heuristic takes the idea a little further. 

Given a candidate tuple $t$ and one of its beyond-$h$ queries $q$, there are two possible results : (1) $t$ is the top ranked non-top-$h$ tuple (2) $t$ is {\em not} the top ranked non-top-$h$ tuple. In the first case, the query $q$ did not give any contradicting evidence for $t$ and the next beyond-$h$ query needs to be executed. On the other hand, the second outcome provides an evidence that disqualifies $t$ from being the next ranked tuple. i.e. the procedure for testing $t$ can be terminated early.  The heuristic tries to reorder the execution of beyond-$h$ queries so that if $t$ is not the No. $h+1$ ranked tuple, it is detected earlier.

While reordering the queries of a single candidate tuple is useful by itself, the maximum advantage is obtained when the set of beyond-$h$ queries of {\em all} the tuples in candidate set are reordered. By ordering queries based on the chance that it eliminates atleast one candidate tuple and executing them in that order, we eliminate as many candidates as possible in the least number of queries. Furthermore, while executing the queries, any candidate tuple dominate by others can be immediately rejected. 

The heuristic relies on two factors that make a beyond-$h$ query $q$ useful. Note that both the factors implicitly favor shorter queries over longer ones. 
\begin{itemize}
\item The number of tuples in {\em candidate set} matched by $q$. If $q$ matches $l$ tuples in candidate set, we can immediately eliminate the $l-1$ dominated candidates after executing $q$ as they cannot have rank $h+1$.
\item The {\em expected} number of tuples in the {\em database} that is matched by $q$. If $q$ matches a large fraction of database, then there is a high likelihood that one of such tuples will be ranked higher than candidate tuple $t$. Of course, since the entire database is not available to us, we estimate the fraction by assuming a random database where the attribute values are uniformly distributed. While this assumption does not always hold, it serves as a useful approximation and heuristic. Given a boolean database with $5$ attributes and any query with two attributes can be expected to match 25\% of the tuples.
\end{itemize}

In summary, the query ordering heuristic pools the beyond-$h$ queries of all candidate tuples and reorders them based on a weighted combination of the two factors described above. The weights can be determined using domain knowledge of the hidden database. The queries are executed in the order so as to eliminate the candidate tuples as early as possible. Any candidate tuple dominated by a non top-$h$ tuple or other candidate tuple are eliminated. The process is continued till only one candidate remains.

\textbf{Example :} Suppose we wanted to determine if $t_3$ or $t_4$ is the third ranked tuple. $A3=1$ and $A4=1$ are two of the beyond-$h$ queries for $t_4$ while the corresponding ones for $t_3$ are $A3=1$ and $A4=0$. Since the query $A3=1$ matches both $t_3$ and $t_4$, it is executed before either of $A4=1$ or $A4=0$. After executing $A3=1$, we note that $t_3$ is ranked higher than $t_4$ in the result and hence declare it as the 3-rd ranked tuple.

\textbf{Analysis : } The query cost of heuristic is bounded by the upper bound for the number of beyond-$h$ queries for the tuples in candidate set. In the worst case, this procedure degenerates to executing all the beyond-$h$ queries for {\em all but one} of the candidate tuples.

\section{Algorithm Design and Extensions}
\label{sec:extensions}
In this section, we integrate the candidate generation and testing techniques discussed in previous two sections to develop our final algorithms for \textsc{GetNext}. In addition, we shall describe different extensions of our algorithms such as retrieving the top ranked tuples when no unique total order exists among them or retrieving top ranked tuples that satisfy additional user specified filters.

\subsection{Algorithms BEYOND-h-GETNEXT and\\ ORDERED-GETNEXT}

We start by integrating our DAG-based candidate generation algorithm with the beyond-$h$ queries based candidate testing algorithm to develop the BEYOND-$h$-GETNEXT algorithm. To be the next ranked tuple, any candidate tuple must be the top ranked non top-$h$ tuple for each of its beyond-$h$ queries. Algorithm~\ref{alg:beyondHFinder} depicts the pseudocode of BEYOND-$h$-GETNEXT.

% Algorithm : BEYOND-$h$-GETNEXT
\begin{algorithm}[!htb]
\caption{\textbf{ BEYOND-$h$-GETNEXT}}
\begin{algorithmic}[1]
\label{alg:beyondHFinder}
\STATE \textbf{Input parameters :} $topH$, the set of top ranked tuples
\STATE Get candidates for $t_{h+1}$ using candidate generation
\FOR {each candidate tuple $t$ }
\STATE Generate and execute beyond-$h$-queries for $t$
\STATE If any tuple other than top-$h$ tuples dominate $t$, reject $t$
\ENDFOR
\RETURN unrejected tuple as $t_{h+1}$
\end{algorithmic}
\end{algorithm}

We also integrate our candidate generation algorithm with the heuristic candidate testing algorithm to develop the ORDERED-GETNEXT algorithm. The only difference between ORDERED-GETNEXT and BEYOND-$h$-GETNEXT is in the rank testing phase. In ORDERED-GETNEXT, we first identify the beyond-$h$ queries for {\em all} candidate tuples and order them based on their likelihood of rejecting a candidate tuple. The queries are executed until all but one candidate tuples have been rejected. The remaining tuple is declared as No.~$h+1$ tuple. Algorithm ~\ref{alg:orderedFinder} depicts the pseudocode of ORDERED-GETNEXT.

% Algorithm : ORDERED-GETNEXT
\begin{algorithm}[!htb]
\caption{\textbf{ ORDERED-GETNEXT}}
\begin{algorithmic}[1]
\label{alg:orderedFinder}
\STATE \textbf{Input parameters :} $topH$, the set of top ranked tuples
\STATE Get candidates for $t_{h+1}$ using candidate generation
\STATE Collect the beyond-$h$ queries of all candidates and order them based on likelihood to reject candidates
\FOR {each query}
\STATE Execute query
\STATE Reject any candidate tuple dominated by other candidates or a non top-$h$ tuple
\STATE If only one candidate left, break
\ENDFOR
\RETURN remaining tuple as $t_{h+1}$
\end{algorithmic}
\end{algorithm}

% Query results with partial order
\subsection{Absence of Total Order within Top Ranked Tuples}
\label{subsec:partialOrder}

One of the assumptions that was made by the algorithms was that the set of top ranked tuples that we wish to retrieve are totally ordered and the order is inferable from the hidden database interface. Specifically, we assumed that tuples $t_h$ and $t_{h+1}$ was directly comparable. In this subsection, we discuss how to handle the different scenarios when the assumption does not hold. 

Two tuples can be compared with each other either directly or indirectly and similarly the dominance relationship can be established directly or indirectly through other intermediate tuples. For eg, we might have two tuples $t$ and $v$ that are not directly comparable. However, if $t \succ u$ and $u \succ v$, then we can indirectly infer their dominance relationship.  If two tuples are not comparable at all, even indirectly, then their dominance relationship cannot be established. Choosing either of the tuples to be the next ranked tuple results in a potentially valid total ordering from the limited information available. The possibility of two tuples not comparable affects both the candidate generation and testing steps.

\textbf{Candidate Generation:} In candidate generation, if the head tuple of {\em every} linear chain was not comparable to $t_h$, then we cannot assign the head tuple from the linear chain from which $t_h$ was extracted to be the next ranked tuple (as it is not comparable to $t_h$). All the non dominated candidate tuples are sent to candidate testing for identifying the next ranked tuple.

\textbf{Candidate Testing:} If multiple tuples from candidate set are not dominated by any other tuple other than the ones in top-$h$ (including other tuples in candidate set), then each of them can potentially be considered as the next ranked tuple. Hence, one of the non dominated candidate tuples is selected uniformly at random as the next ranked tuple and the process is continued. This random selection creates one of the valid partial order of the top ranked tuples. Since the output total order is no longer accurate, a metric must be chosen to measure the distance between the actual total order and the partial order. The accuracy measure used is the expected distance between a randomly generated total order and the actual total order. The distance between two ranked list can be computed using Kendall $\tau$ or the Spearman's footrule.

% Queries with selection constraint
\subsection{Top Ranked Tuples with Selectivity Constraints}
The discussions in the previous sections described techniques to retrieve the top ranked tuples from the entire database. An equally important and practical scenario is one where the user is interested in the top ranked tuples over a subset of the database. For example, the user might be interested in the cheapest flights with in-flight wifi. An alternate perspective is to view the problem as retrieving top ranked tuples where some of the attribute values are already preset by the user, for e.g. wifi. The specified attributes then partition the entire hidden database into two partitions - one which matches the specified attributes and another which does not match the specified attributes. In this subsection, we discuss how to extend the techniques discussed so far to solve this problem.

An initial approach one might come up with is to keep retrieving top tuples from the {\em entire} database incrementally till we have adequate number of tuples satisfying the user selectivity constraints. This might be the only possible approach if the user selectivity constraint cannot be filtered through the interface of hidden database. For e.g. if the user is interested in top-10 flights with in-flight wifi. However if the constraint cannot be entered via the airline interface, we can keep retrieving top ranked tuples till we have accumulated 10 flights with in-flight wifi. If the filters are too selective, then the number of tuples to be fetched before we return the user results could be very high.

However, if the user's constraints can be entered via the hidden database interface (but user still needs more that $k$ results), then an alternate approach is possible. As an example, the user might be interested in top-20 flights with wifi on a top-10 interface where the wifi availability is an input attribute. We can directly apply the techniques for extracting top ranked tuples over the subset of database that satisfies the selectivity constraints instead of applying it on the original database. This corresponds to prefixing the selectivity constraints to each of the queries executed by the algorithms. The candidate generation phase produces only tuples that satisfy the constraints.

The algorithms that work only on the database subset might seem to be a more efficient approach to solve the problem and in most scenarios it is. However, there are few factors that influence the output. First, if the selectivity constraints are coarse or not too selective, then a large section of database would be covered. This in turn, increases the chances of finding a correct set of top ranked tuples satisfying the constraints. If the number of tuples that match are small, then there is a high likelihood that the tuples are not comparable. In this case, we are left with a partial order of tuples instead of a total order.

Secondly, even if a total order exist among the top ranked tuples in the subset, it might not be possible to order them by only looking at the candidate tuples matching the constraints. This is because, the tuple(s) that helped to indirectly compare and order the candidate tuples, say $t_x$ and $t_y$ could itself not satisfy the selectivity constraint. In this case, the two tuples are incomparable, even though a global order exist between them. In both the scenarios, we are potentially left with a partial order. The techniques used in linearizing the partial order from $\S$6-B can be used to solve this issue.

\section{Experimental Results}
\label{sec:experiments}

\begin{figure*}[t]
\begin{minipage}[t]{0.23\linewidth}
\centering
\includegraphics[height=45mm, width = 50mm]{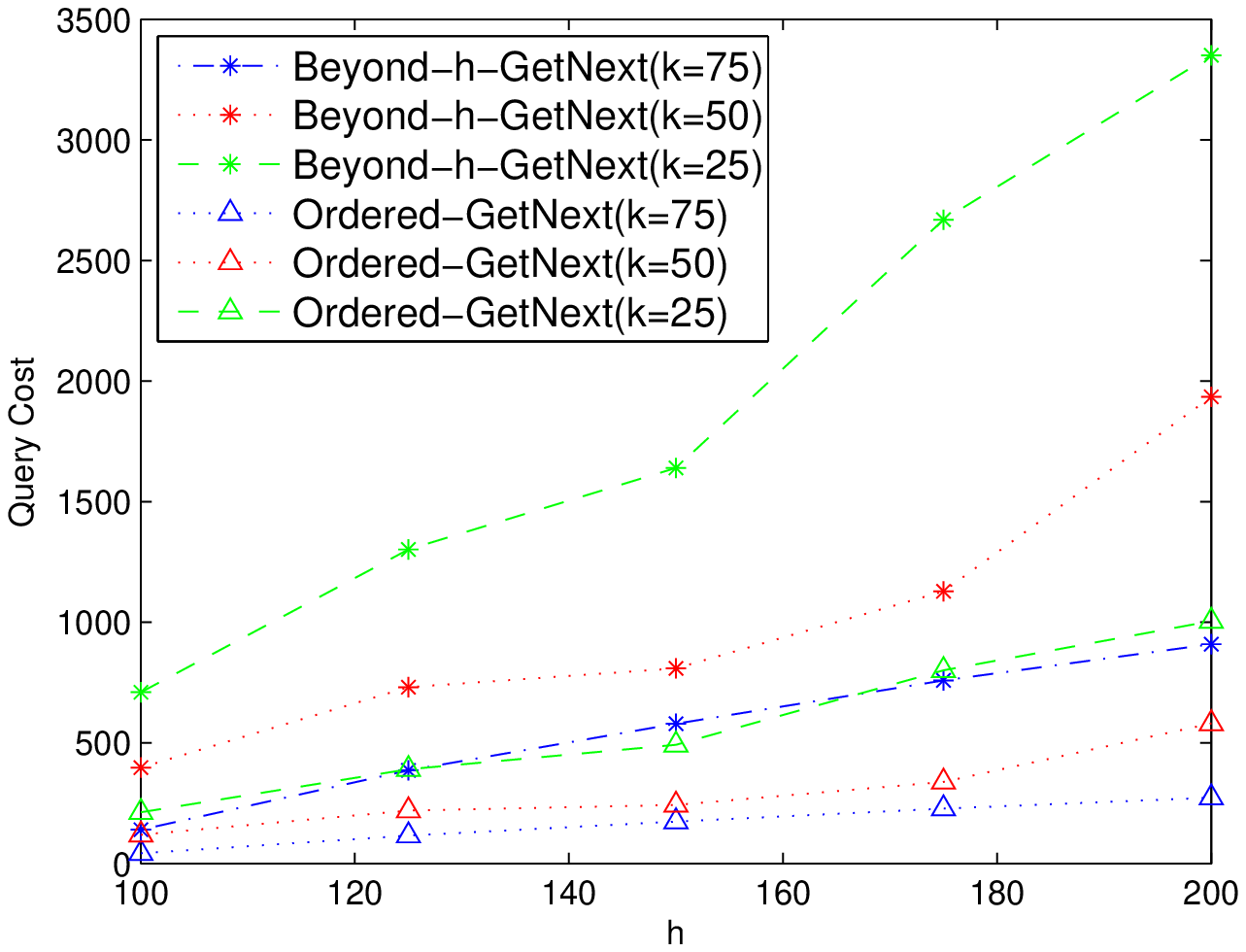}
\caption{Query cost vs $h$ on Boolean dataset}
\label{fig:expt-1}
\end{minipage}
\hspace{3mm}
\begin{minipage}[t]{0.23\linewidth}
\centering
\includegraphics[height=45mm, width = 50mm]{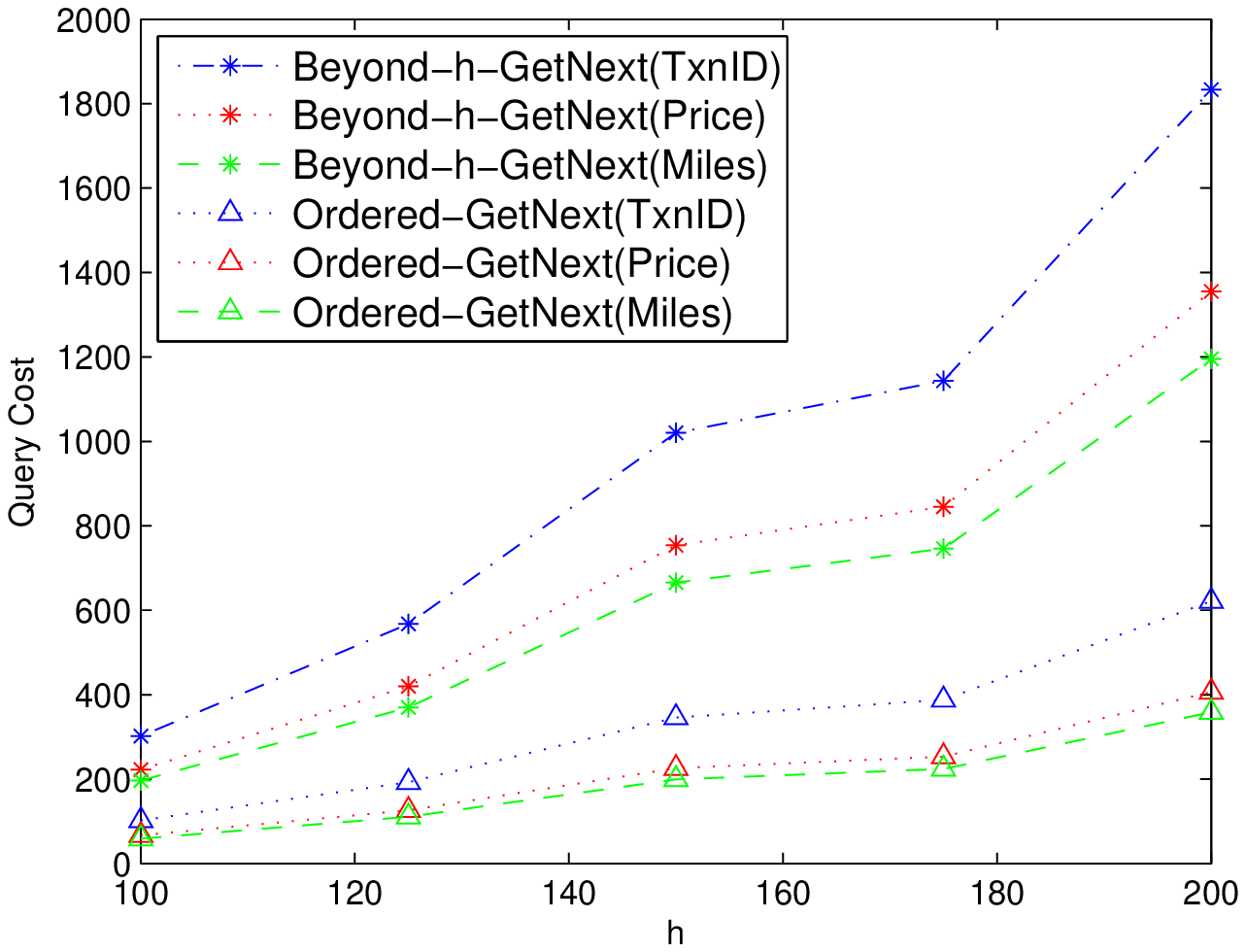}
\caption{Effect of different ranking functions on Autos dataset}
\label{fig:expt-2}
\end{minipage}
\hspace{3mm}
\begin{minipage}[t]{0.23\linewidth}
\centering
\includegraphics[height=45mm, width = 50mm]{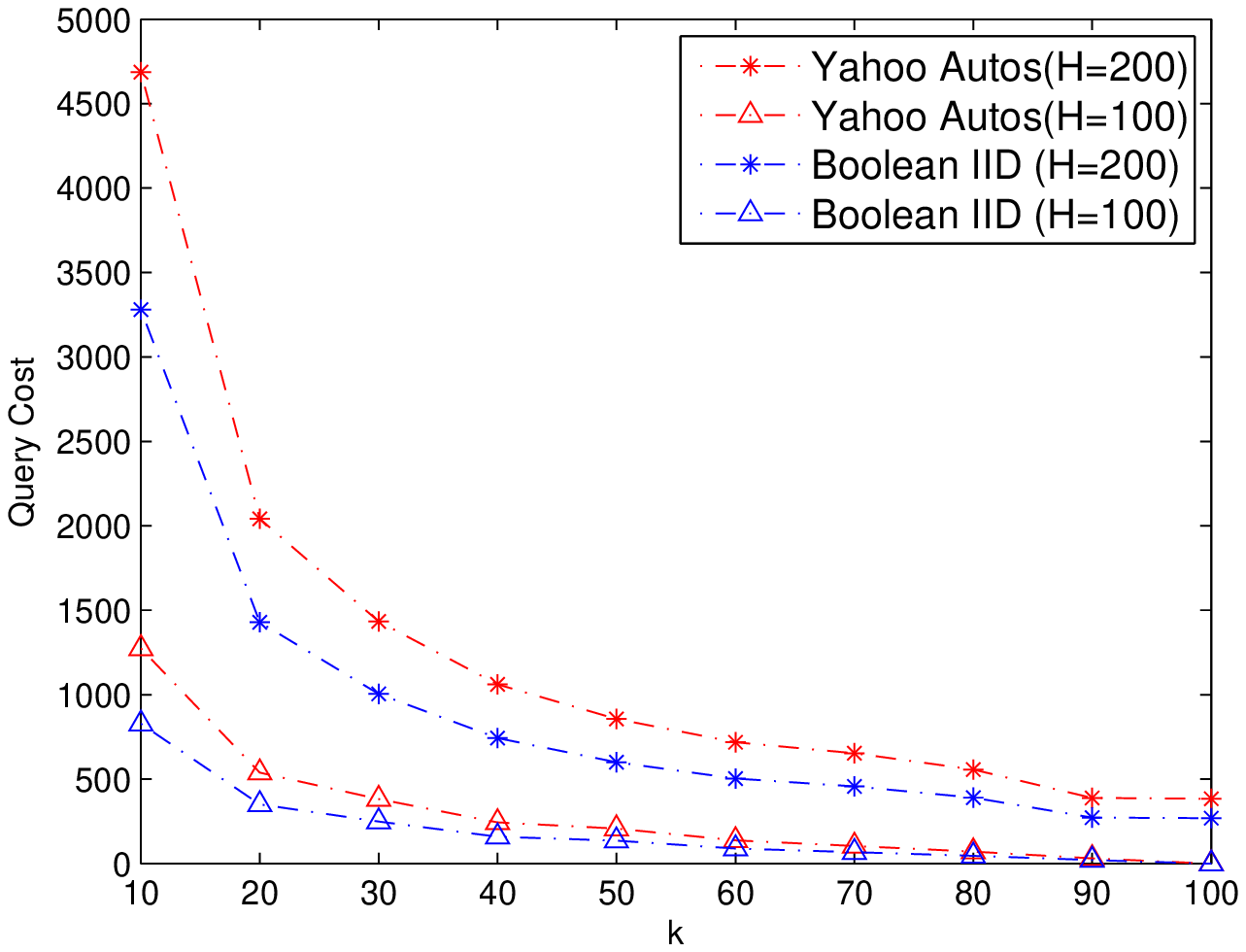}
\caption{Query cost versus $k$}
\label{fig:expt-3}
\end{minipage}
\hspace{3mm}
\begin{minipage}[t]{0.22\linewidth}
\centering
\includegraphics[height=45mm, width = 50mm]{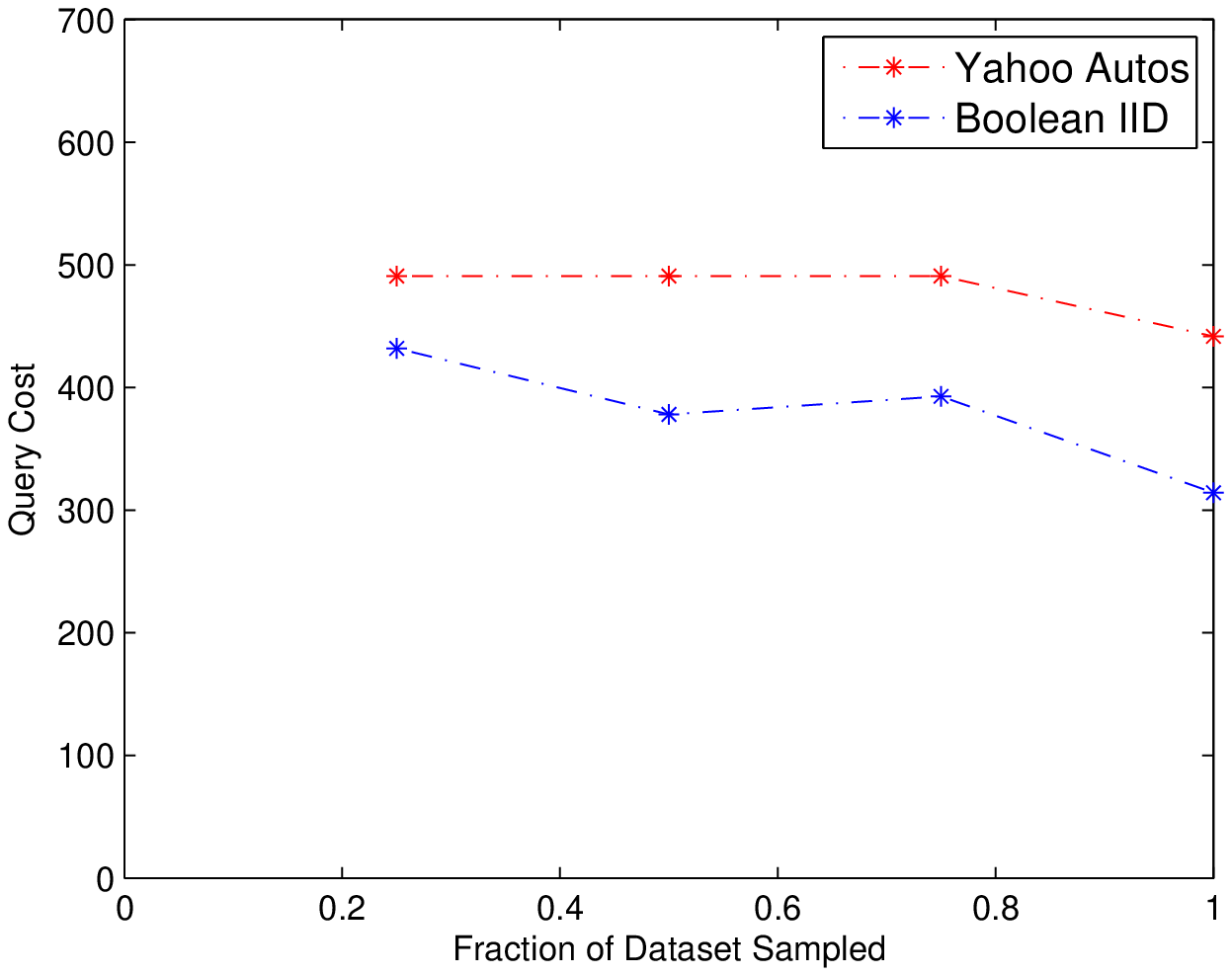}
\caption{Query cost versus database size}
\label{fig:expt-4}
\end{minipage}
\hspace{-2mm}
\end{figure*}

%line 2 
\begin{figure*}[t]
\begin{minipage}[t]{0.23\linewidth}
\centering
\includegraphics[height=45mm, width = 50mm]{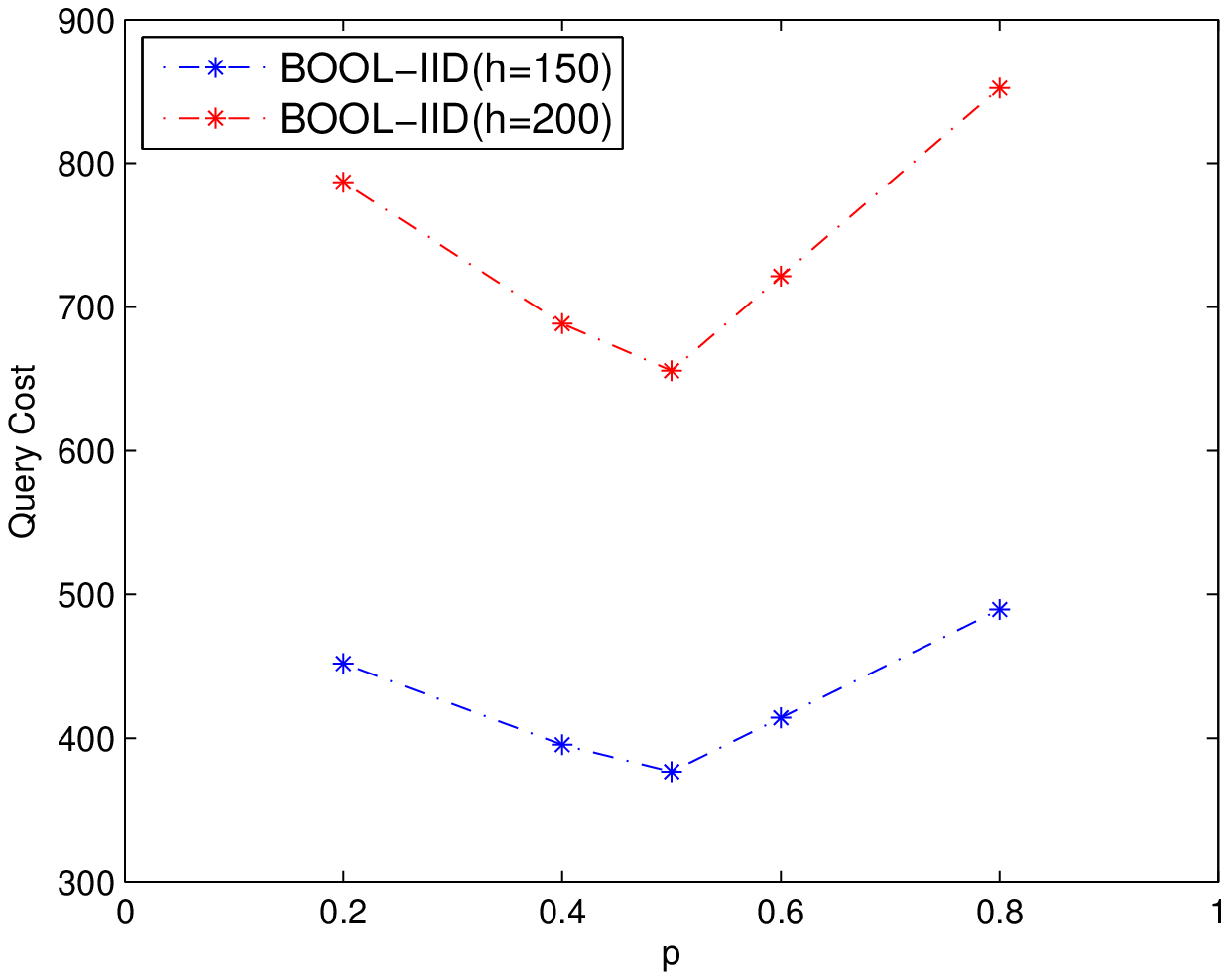}
\caption{Query cost versus skew}
\label{fig:expt-5}
\end{minipage}
\hspace{3mm}
\begin{minipage}[t]{0.23\linewidth}
\centering
\includegraphics[height=45mm, width = 50mm]{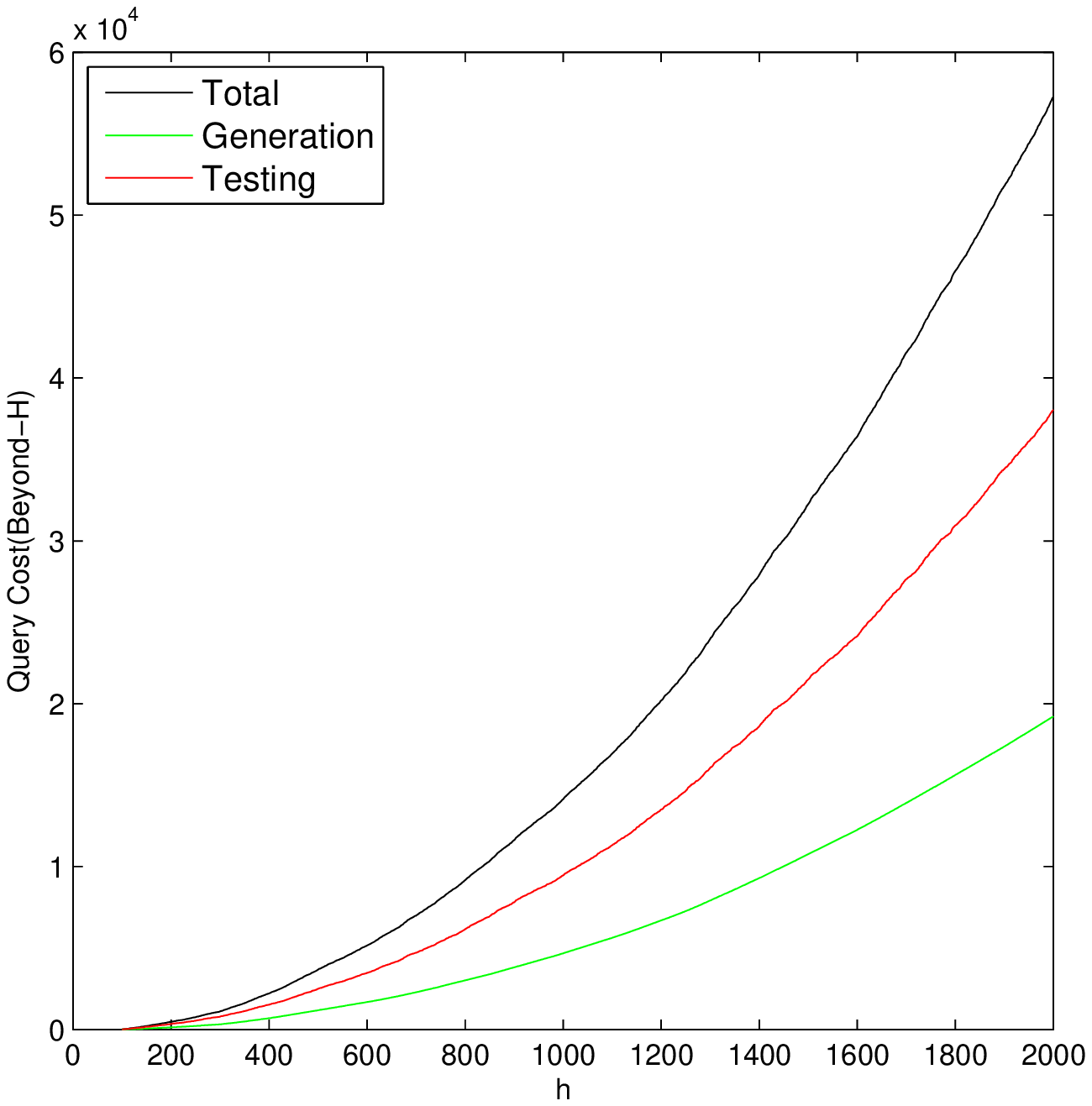}
\caption{Query cost versus large $h$}
\label{fig:expt-6}
\end{minipage}
\hspace{3mm}
\begin{minipage}[t]{0.23\linewidth}
\centering
\includegraphics[height=45mm, width = 50mm]{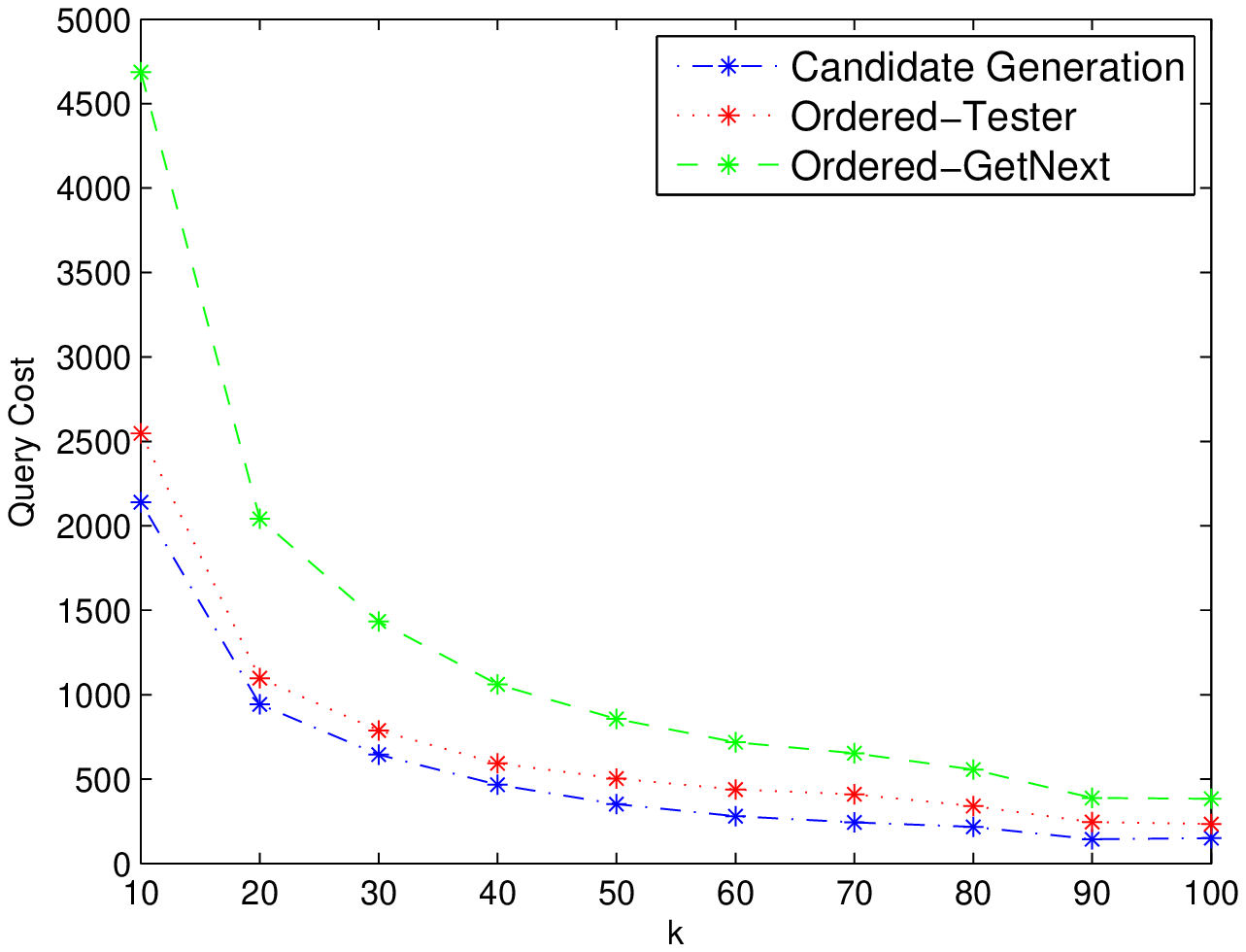}
\caption{Comparing candidate generation versus testing}
\label{fig:expt-7}
\end{minipage}
\hspace{3mm}
\begin{minipage}[t]{0.23\linewidth}
\centering
\includegraphics[height=45mm, width = 50mm]{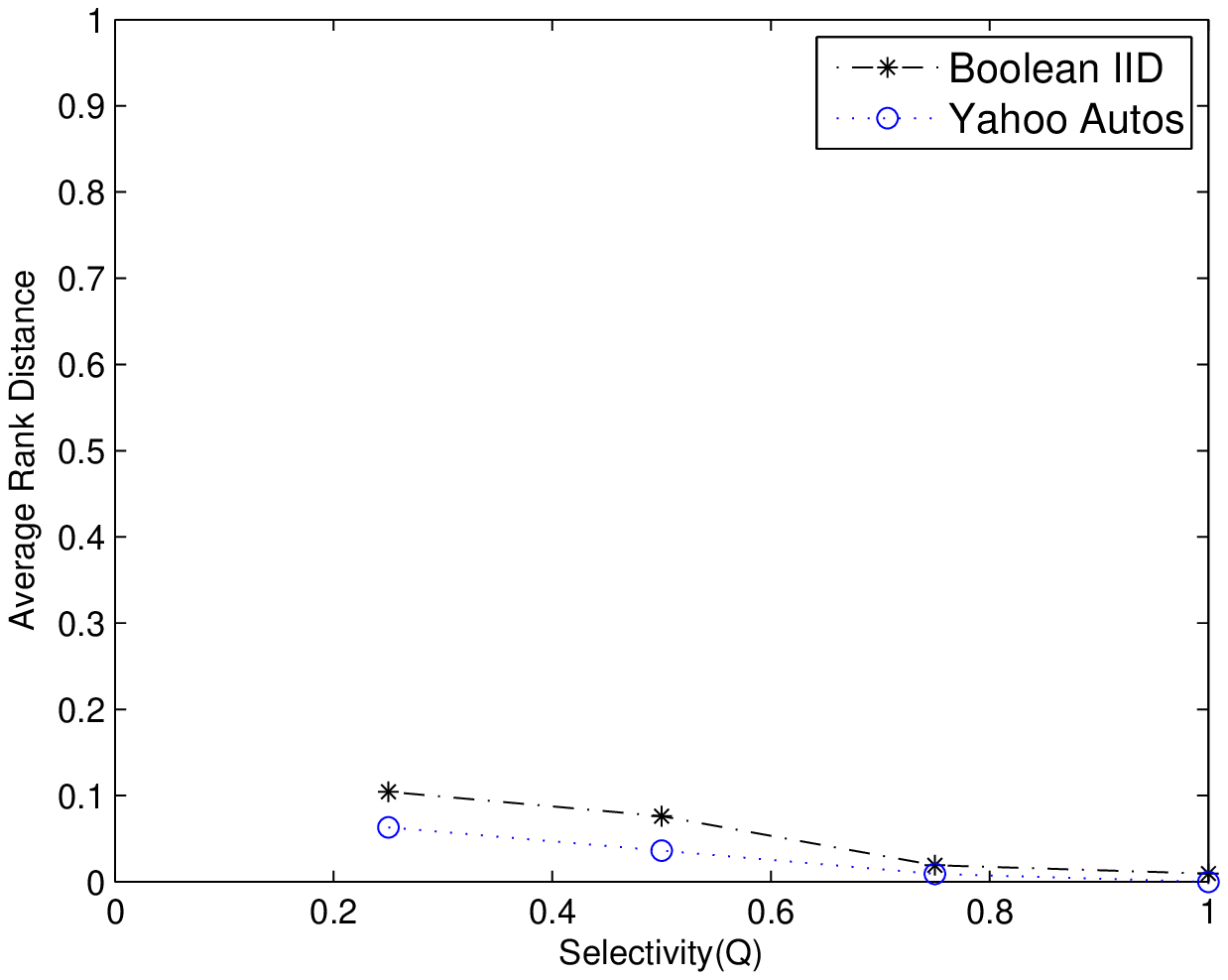}
\caption{Rank distance versus query selectivity}
\label{fig:expt-8}
\end{minipage}
\hspace{-5mm}
\end{figure*}

In this section we describe our experimental setup, compare the performance of algorithms for candidate generation and candidate testing and show the efficiency and accuracy of our methods.

\subsection{Experimental Setup}

\smallskip\noindent
{\bf Hardware and Platform:} All our experiments were performed on a quad-core 2 GHz AMD Phenom machine with 8 GB of RAM. The algorithms were implemented in Python.

\smallskip\noindent
{\bf  Datasets:} We used both synthetic and real-world data sets in the experiments. The synthetic dataset we used is a boolean one with 200,000 tuples and 80 attributes. The tuples are generated as i.i.d. data with each attribute having probability of $p$ = 0.5 to be 1 (except for one experiment where we created different datasets with different values of $p$). We refer to this dataset as the BOOL-IID dataset. The real-world dataset we used consists of data crawled from the Yahoo! Autos website \footnote{\url{http://autos.yahoo.com/}}, a real-world hidden database. It contains 200,000 used cars for sale in the Dallas-Fort Worth metropolitan area. There are 32 Boolean attributes such as A/C, Power Locks, etc, and 6 categorical attributes, such as Make, Model, Color, etc. The domain size of categorical attributes ranges from 5 to 16.

\smallskip\noindent
{\bf  Real-World Online Experiment:} In addition to the offline experiments described above, we also directly applied our techniques {\em online} over Amazon.com (specifically Amazon's Product Advertising API\footnote{https://affiliate-program.amazon.com/gp/advertising/api/detail/main.html}) to discover the top-250 (according to sales rank) Amazon DVD titles from a top-100 interface\footnote{By default Amazon's Product Advertising API provides a top-10 interface, while allowing a user to ``Page Down" for up to 9 times, essentially leading to a top-100 interface.} provided by the API. Since the individual item description provided by Amazon.com reveals the sales rank of the item, we were able to verify the correctness of all results discovered by our algorithm. For this online experiment, (top-$k$) search query can be constructed using 15 categorical attributes such as Actor, Artist, Publisher, etc., with their domain sizes ranging from 5 to over 1,000. Amazon.com has a limit of 2,000 queries per IP address per hour.

\smallskip\noindent
{\bf  Algorithms:} We tested two algorithms BEYOND-$h$-GETNEXT and ORDERED-GETNEXT. However, since both these algorithms use the same candidate generation technique, we highlight the behavior of the candidate generation and testing phase separately. In other words, we plot the performance of algorithm GETNEXT for different parameters and then compare the performance of different candidate testing algorithms. This choice of presentation accentuates the improvements provided by the beyond-$h$-queries and the heuristic query ordering that gets masked when directly comparing BEYOND-$h$-GETNEXT and ORDERED-GETNEXT.

\smallskip\noindent
{\bf   Performance Measures:} We use query cost, the number of queries executed on the hidden database as the performance measure. This includes the queries used to retrieve candidate tuples, queries to compare candidates and the beyond-$h$ queries for each candidate. When the total order cannot be inferred, we use expected distance between randomly generated total order and the actual total order. The distance between two ranked lists is computed using Kendall-$\tau$ metric.

\subsection{Experimental Results}

%figure 1 - x axis: h, y axis: query cost, three lines, k = 25, 50, 75 for Boolean IID. Both BEYOND-k-GETNEXT and ORDERED-GETNEXT
%figure 2 - x axis: h, y axis: query cost, three lines, three ranking functions for Yahoo! Autos. Both BEYOND-k-GETNEXT and ORDERED-GETNEXT

In the following discussion we denote the number of top ranked tuples retrieved from the hidden database as $h$. In other words, it denotes the maximum number of invocations of GETNEXT by the third party service.

\smallskip\noindent
{\bf   Query cost versus $h$:} In our first experiment, we evaluated the performance of our algorithms BEYOND-$h$-GETNEXT and ORDERED-GETNEXT on the boolean dataset by investigating the query cost as a function of $h$ for various different values
of $k$. As Figure~\ref{fig:expt-1} shows, the query cost increases with increasing $h$, as is expected. Moreover, significant
savings are achieved by using the ordering heuristic in ORDERED-GETNEXT. We also notice that $k$ plays an important role in the
efficiency of the algorithms: larger $k$ results in more efficient performance. To consider a specific performance
point, when $k = 75$ and $h = 200$, ORDERED-GETNEXT requires less than 300 additional queries to retrieve the extra 125 tuples.

We also performed similar experiments on the Autos dataset and observed similar trends, with ORDERED-GETNEXT outperforming
BEYOND-$h$-GETNEXT (Figure ~\ref{fig:expt-2}). Additionally, we also investigated the
effect the specific ranking function used has on the performance of our algorithms. As Figure ~\ref{fig:expt-2} shows, we used
three different ranking attributes: TxnID (a unique ID for each tuple), as well as attributes such as Price and Miles.
We note that the performance of our algorithms vary for different ranking functions, but nevertheless are still very efficient
in all cases (and as noted earlier, our algorithms do not try to take advantage of any knowledge of these ranking functions).

%figure 3 - x axis: k, y axis: query cost, four lines, two for Boolean IID (h=100 and 200), two for Yahoo! Autos (h=100 and 200), fixed k (say 100)
\smallskip\noindent
{\bf   Query cost versus $k$:} In our next experiment, we investigated the effect of $k$ on the query cost for fixed values of $h$, for
both the boolean dataset as well as the Autos dataset. As
Figure ~\ref{fig:expt-3} shows, the positive effect of larger values of $k$ on the query cost is dramatic, with larger values of $k$
being very effective in reducing the query cost of our algorithms. This is to be expected, as our earlier arguments in the
paper have shown that large $k$ significantly reduces the number of queries needed in the candidate generation and testing procedures
(since the number of minimal infrequent itemsets in a database rapidly reduces with increasing support threshold).

%figure 4 - x axis: db size, y axis: query cost, two lines, one each for Boolean IID and Yahoo! Autos, fixed h = 200, fixed k = 100
\smallskip\noindent
{\bf   Query cost versus database size:} Since our algorithms are designed to retrieve only the top-$h$ tuples from the database, the actual size of the database should not have a significant impact on the performance of our algorithms. This is verified in Figure
~\ref{fig:expt-4}, which shows that the query cost remained practically unchanged for ORDERED-GETNEXT, even though we try our experiments on various fractional sizes of the original databases (the slight dip in query cost is attributable to the uncertainty of
the sampling process). In this experiment, $k = 100$ and $h = 200$.

%figure 5 - x axis: p used in generating Bool-IID, y axis: query cost, three lines, two lines with h = 150 and 200, respectively. fixed k = 100
\smallskip\noindent
{\bf   Query cost versus skew:} We experimented with ORDERED-GETNEXT ($k=100$, $h=200$) on several boolean databases created with different
values of skew parameter $p$. As Figure
~\ref{fig:expt-4} shows, the algorithm is most efficient when the database has equiprobable 1s and 0s, but the cost increases when the proportion becomes unbalanced. This is attributable to the fact that when the database contains more 1s (or more 0s), 
the algorithm has to ``dig deeper'' - i.e., issue a larger number of (and more specific) queries in order to generate all candidates.

\begin{figure}[t]
\centering
\includegraphics[height=40mm, width = 75mm]{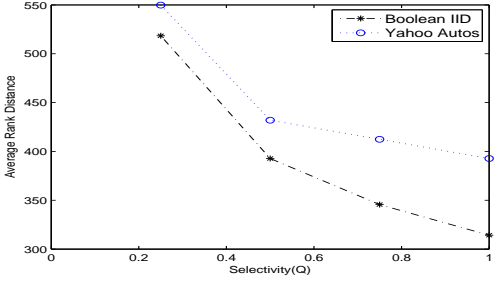}
\caption{Query Cost versus Query Selectivity}
\label{fig:expt-9}
\end{figure}
\begin{figure}[t]
\centering
\includegraphics[height=40mm, width = 75mm]{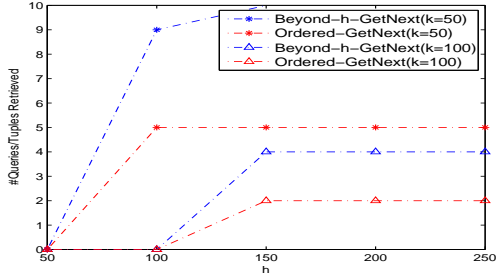}
\caption{Query Cost versus $h$ on Amazon DVD website }
\label{fig:expt-10}
\end{figure}

%figure 6 - x axis: h (long range showcasing the limit of our tech), y axis: query cost, three lines, generation only, testing only, total. over Yahoo! Autos. fixed k = 100.
\smallskip\noindent
{\bf   Effect of large $h$:} Our earlier experiments were focused on values of $h$ that were at most a small factor larger than $k$. Such values are meaningful in actual applications where an user is interested in seeing a few more tuples than what has been returned to her by the original query. But we were also interested in stress-testing our algorithms on much large values of $h$ to see how
they performed. Figure ~\ref{fig:expt-6} shows the results of such an experiment using ORDERED-GETNEXT on the Autos dataset, where $k$ was set at 100. As can be seen, the
query cost increases quite significantly for much larger values of $h$, which leads to the conclusion that beyond a certain point,
it is actually preferable to {\em crawl} the database and extract the top-$h$ queries rather than use ORDERED-GETNEXT. The figure also
profiles the separate query costs of the candidate generation and testing procedures.

%figure 7 - x axis: k, y axis: query cost, three lines, generation only, testing only, total. over Yahoo! Autos. fixed h = 200.
\smallskip\noindent
{\bf   Comparing generation versus testing procedures:} In Figure ~\ref{fig:expt-7}, we compare the query costs of the two main procedures: candidate generation and candidate testing. We ran ORDERED-GETNEXT over the Autos dataset for $h=200$ and varied $k$.
As can be seen, the query cost is almost equally divided between the generation and test procedures for almost all points of the
curve, with testing being slightly more expensive.

%figure 8 - x axis: selection ratio of conditional query, y axis: rank distance, two lines, one each for Boolean IID and Yahoo! Autos. fixed k = 100, fixed h = 200
%figure 9 - x axis: selection ratio of conditional query, y axis: query cost, two lines, one each for Boolean IID and Yahoo! Autos. fixed k = 100, fixed h = 200
\smallskip\noindent
{\bf   Effect of query selectivity:} In Figure~\ref{fig:expt-7}, we investigate the relation between query cost and selectivity. If a query is extremely selective, then it is clear that no algorithm can extract a total order of the top-$h$ tuples. In such situations,
our algorithms return a partial order of the top-$h$ tuples. In this experiment, we compare a random total order that comforms to
the returned partial order against the true top-$h$ tuples for that query using Kendall-$\tau$ measure. 
As the query becomes less selective, the rank distance increases and its query cost becomes less, which is to be expected as the candidate testing procedure gets opportunities to terminate early as one needs a smaller number of queries to exclude a tuple from consideration.
Our experiments uses ORDERED-GETNEXT for both datasets, with $k=100$ and $h=200$.

\smallskip\noindent
{\bf Experiment against Amazon DVD Titles :} To show the practicality of our algorithms, we retrieved the top-250 Amazon DVD titles in terms of their sales rank. Note that by default, Amazon only displays the top-100 items in any category. The correctness of our algorithm is verified by the checking the individual item description pages of the items discovered by GETNEXT (which reveals the actual sales ranking of the items). The queries were made using the Amazon Product Advertising API and the maximum value of $k$ is 100. A sample query to get the top-10 PG rated DVDs ordered by their salesrank is shown in footnote\footnote{http://ecs.amazonaws.com/onca/xml?Service=AWSECommerceService\\\&AWSAccessKeyId=[fill]\&Operation=ItemSearch\&SearchIndex=DVD\\\&ResponseGroup=Large,SalesRank\&Sort=salesrank\&AudienceRating=PG\\\&Timestamp=[fill]\&Signature=[fill]}.Figure~\ref{fig:expt-10} shows that when $k=100$, the top-250 titles can be retrieved using fewer that 500 queries, well below the 2000 queries-per-hour-per-IP-address limit imposed by Amazon.com. The figure also shows the behavior of both BEYOND-$h$-GETNEXT and ORDERED-GETNEXT for different values of $k$ and $h$.

\section{Related Work}
\label{sec:relatedWork}

\noindent {\bf Information Integration and Extraction for Hidden databases:} A significant body of research has been done on information integration and extraction over hidden databases - see tutorials \cite{CC:06, DRV:06}. Due to space limit, we only list a few closely-related work: \cite{RG01} proposes a crawling solution. Parsing and understanding web query interfaces has been extensively studied (e.g., \cite{DKYL:09, ZHC:04}). The mapping of attributes across different web interfaces has also been addressed (e.g., \cite{HCH:04}). Also related is the work on integrating query interfaces for multiple web databases in the same topic-area (e.g., \cite{DYM:06, HC:03}). Our paper provides results orthogonal to these existing techniques as it represents the first formal study on retrieving top-$h$ ($h > k$) tuples matching a user-specified query by reformulating the query through a top-$k$ interface.

\noindent {\bf Data Analytics over Hidden Databases:} There has been prior work on crawling, sampling, and aggregate estimation over the hidden web, specifically over text \cite{BG:08b, BB:98} and structured \cite{RG01} hidden databases and search engines \cite{LYM:02, SZS+:06, BG:07}. Specifically, sampling-based methods were used for generating content summaries \cite{CC:01, IG:02, HYJS:06}, processing top-$k$ queries \cite{BGM:02}, etc. Prior work (see \cite{DJJ+10} and references therein) considered sampling and aggregate estimation over structured hidden databases.

\noindent {\bf Top-$k$ Query Processing:} There have been extensive studies on retrieving the top-$k$ tuples over a traditional database - see \cite{IBS08} for a survey. Our approach differs by allowing the retrieval of top-$h$ tuples through a restricted top-$k$ web interface.

\noindent {\bf Frequent Itemset Mining:} In $\S$5, we map the discovery of beyond-$h$ queries to the problem of infrequent-minimal-itemset mining - a problem well studied in data mining \cite{HCXY07}. \cite{HM07} provides additional details about algorithms and properties for infrequent itemset mining.

\section{Conclusion}
\label{sec:conclusion}

In this paper we have initiated study on the problem of retrieving the top-$h$ ($h>k$) tuples from a hidden web database that only provides a top-$k$ search interface. To address the fundamental operator \textsc{GetNext}, we proposed a two-step process, candidate generation and candidate testing, and developed efficient algorithms for both steps. We conducted comprehensive set of experiments over synthetic datasets and real-world hidden databases which demonstrate the effectiveness of our proposed techniques. There are multiple exciting directions for future research. We intend to investigate the possibility of retrieving the top ranked tuples approximately - for e.g., retrieve as many top ranked tuples under budget cost or in a rank agnostic fashion. Further, we plan to build attractive demonstrations of mashup applications against real-world hidden web databases.

% ensure same length columns on last page (might need two sub-sequent latex runs)
%\balance

% The following two commands are all you need in the
% initial runs of your .tex file to
% produce the bibliography for the citations in your paper.
\bibliographystyle{IEEEtran}
\bibliography{IEEEabrv,breakingTopKBarrier}  % 

\begin{thebibliography}{10}
\providecommand{\url}[1]{#1}
\csname url@rmstyle\endcsname
\providecommand{\newblock}{\relax}
\providecommand{\bibinfo}[2]{#2}
\providecommand\BIBentrySTDinterwordspacing{\spaceskip=0pt\relax}
\providecommand\BIBentryALTinterwordstretchfactor{4}
\providecommand\BIBentryALTinterwordspacing{\spaceskip=\fontdimen2\font plus
\BIBentryALTinterwordstretchfactor\fontdimen3\font minus
  \fontdimen4\font\relax}
\providecommand\BIBforeignlanguage[2]{{%
\expandafter\ifx\csname l@#1\endcsname\relax
\typeout{** WARNING: IEEEtran.bst: No hyphenation pattern has been}%
\typeout{** loaded for the language `#1'. Using the pattern for}%
\typeout{** the default language instead.}%
\else
\language=\csname l@#1\endcsname
\fi
#2}}

\bibitem{MKK+08}
J.~Madhavan, D.~Ko, L.~Kot, V.~Ganapathy, A.~Rasmussen, and A.~Y. Halevy,
  ``{Google's Deep Web crawl},'' \emph{Proceedings of The Vldb Endowment},
  vol.~1, pp. 1241--1252, 2008.

\bibitem{ARJ+07}
M.~\'{A}lvarez, J.~Raposo, A.~Pan, F.~Cacheda, F.~Bellas, and V.~Carneiro,
  ``Crawling the content hidden behind web forms,'' in \emph{Proceedings of the
  2007 international conference on Computational science and Its applications -
  Volume Part II}, ser. ICCSA'07.\hskip 1em plus 0.5em minus 0.4em\relax
  Springer-Verlag, 2007, pp. 322--333.

\bibitem{DJJ+10}
A.~Dasgupta, X.~Jin, B.~Jewell, N.~Zhang, and G.~Das, ``Unbiased estimation of
  size and other aggregates over hidden web databases,'' in \emph{SIGMOD},
  2010.

\bibitem{DDM07}
A.~Dasgupta, G.~Das, and H.~Mannila, ``A random walk approach to sampling
  hidden databases,'' in \emph{SIGMOD}, 2007.

\bibitem{JZD11}
X.~Jin, N.~Zhang, and G.~Das, ``Attribute domain discovery for hidden web
  databases,'' in \emph{SIGMOD}, 2011.

\bibitem{HCXY07}
J.~Han, H.~Cheng, D.~Xin, and X.~Yan, ``Frequent pattern mining: current status
  and future directions,'' \emph{DMKD}, 2007.

\bibitem{CC:06}
K.~Chang and J.~Cho, ``Accessing the web: From search to integration,'' in
  \emph{Tutorial, SIGMOD}, 2006.

\bibitem{DRV:06}
A.~Doan, R.~Ramakrishnan, and S.~Vaithyanathan, ``Managing information
  extraction,'' in \emph{Tutorial, SIGMOD}, 2006.

\bibitem{RG01}
S.~Raghavan and H.~Garcia-Molina, ``Crawling the hidden web,'' in \emph{VLDB},
  2001.

\bibitem{DKYL:09}
E.~Dragut, T.~Kabisch, C.~Yu, and U.~Leser, ``A hierarchical approach to model
  web query interfaces for web source integration,'' in \emph{VLDB}, 2009.

\bibitem{ZHC:04}
Z.~Zhang, B.~He, and K.~Chang, ``Understanding web query interfaces:
  best-effort parsing with hidden syntax,'' in \emph{SIGMOD}, 2004.

\bibitem{HCH:04}
B.~He, K.~Chang, and J.~Han, ``Discovering complex matchings across web query
  interfaces: A correlation mining approach,'' in \emph{KDD}, 2004.

\bibitem{DYM:06}
E.~Dragut, C.~Yu, and W.~Meng, ``Meaningful labeling of integrated query
  interfaces,'' in \emph{VLDB}, 2006.

\bibitem{HC:03}
B.~He and K.~Chang, ``Statistical schema matching across web query
  interfaces,'' in \emph{SIGMOD}, 2003.

\bibitem{BG:08b}
Z.~Bar-Yossef and M.~Gurevich, ``Mining search engine query logs via suggestion
  sampling,'' in \emph{VLDB}, 2008.

\bibitem{BB:98}
K.~Bharat and A.~Broder, ``A technique for measuring the relative size and
  overlap of public web search engines,'' in \emph{WWW}, 1998.

\bibitem{LYM:02}
K.~Liu, C.~Yu, and W.~Meng, ``Discovering the representative of a search
  engine,'' in \emph{CIKM}, 2002.

\bibitem{SZS+:06}
M.~Shokouhi, J.~Zobel, F.~Scholer, and S.~Tahaghoghi, ``Capturing collection
  size for distributed non-cooperative retrieval,'' in \emph{SIGIR}, 2006.

\bibitem{BG:07}
Z.~Bar-Yossef and M.~Gurevich, ``Efficient search engine measurements,'' in
  \emph{WWW}, 2007.

\bibitem{CC:01}
J.~Callan and M.~Connell, ``Query-based sampling of text databases,'' \emph{ACM
  TOIS}, vol.~19, no.~2, pp. 97--130, 2001.

\bibitem{IG:02}
P.~Ipeirotis and L.~Gravano, ``Distributed search over the hidden web:
  Hierarchical database sampling and selection,'' in \emph{VLDB}, 2002.

\bibitem{HYJS:06}
Y.-L. Hedley, M.~Younas, A.~E. James, and M.~Sanderson, ``Sampling, information
  extraction and summarisation of hidden web databases,'' \emph{Data and
  Knowledge Engineering}, vol.~59, no.~2, pp. 213--230, 2006.

\bibitem{BGM:02}
N.~Bruno, L.~Gravano, and A.~Marian, ``Evaluating top-k queries over
  web-accessible databases,'' in \emph{ICDE}, 2002.

\bibitem{IBS08}
I.~Ilyas, G.~Beskales, and M.~Soliman, ``A survey of top-k query processing
  techniques in relational database systems,'' \emph{ACM Computing Surveys},
  vol.~40, 2008.

\bibitem{HM07}
D.~J. Haglin and A.~M. Manning, ``On minimal infrequent itemset mining,'' in
  \emph{International Conference on Data Mining}, 2007.

\end{thebibliography}
% You must have a proper ".bib" file
%  and remember to run:
% latex bibtex latex latex
% to resolve all references

\end{document}